\documentclass[10pt, conference,letterpaper]{IEEEtran}

\makeatletter
\def\ps@headings{%
\def\@oddhead{\mbox{}\scriptsize\rightmark \hfil \thepage}%
\def\@evenhead{\scriptsize\thepage \hfil \leftmark\mbox{}}%
\def\@oddfoot{}%
\def\@evenfoot{}}
\makeatother 
\usepackage{graphicx}
\usepackage{epstopdf}
\usepackage{cite}
\usepackage{times}
\usepackage{dsfont} 
\usepackage{cite}
\usepackage{times}
\usepackage{pifont}
\usepackage{multirow}   
\usepackage{tikz}
\usetikzlibrary{shapes.geometric,calc}
   
\usepackage{amsthm}
\usepackage{graphicx}
\usepackage{subfigure}
\usepackage{color} 
\usepackage{ifpdf}
\usepackage{epsfig}
\usepackage{latexsym}
\usepackage{amsfonts}
\usepackage{amssymb}
\usepackage{paralist}
\usepackage{comment}
\usepackage{xspace}
\usepackage{mathrsfs}
\usepackage{amssymb}
\usepackage{setspace}
\usepackage{color}
\usepackage[small]{caption}

\usepackage{subeqnarray}
\usepackage{algorithm}
\usepackage[noend]{algpseudocode}
\usepackage{amssymb}
\usepackage{url,epsfig,array}
\usepackage{leftidx} 
\usepackage{amsmath}
\usepackage[T1]{fontenc}
\usepackage{aecompl}

\usepackage{tikz}

\usepackage[numbers]{natbib}

\usepackage[
top=0.62in, bottom=0.62in,
left=0.6in, right=0.6in
]{geometry} 

\usepackage{pifont}
\usepackage{bm}

\def\ie{\textit{i.e.}\xspace}
\def\etal{\textit{et al.}\xspace}

\def\eg{\textit{e.g.}\xspace}

\makeatletter
\renewcommand{\maketag@@@}[1]{\hbox{\m@th\normalsize\normalfont#1}}%
\makeatother

\begin{document}
  
\title{\Huge{Many Hands Make Light Work: Accelerating Edge Inference via Multi-Client Collaborative Caching}}

\author{\IEEEauthorblockN{Wenyi Liang$^{1,2}$, Jianchun Liu$^{1,2}$, \ Hongli Xu$^{1,2}$,  \ Chunming Qiao$^{3}$, \ Liusheng Huang$^{1,2}$   \\}
\IEEEauthorblockA{  
$^1$School of Computer Science and Technology, University of Science and Technology of China, China\\
$^2$Suzhou Institute for Advanced Research, University of Science and Technology of China, China\\ 
$^3$ Department of Computer Science and Engineering, University at Buffalo, USA\\
}

}

\maketitle

\begin{abstract}
Edge inference is a technology that enables real-time data processing and analysis on clients near the data source.
To ensure compliance with the Service-Level Objectives (SLOs), such as a 30\% latency reduction target, caching is usually adopted to reduce redundant computations in inference tasks on stream data. Due to task and data correlations, sharing cache information among clients can improve the inference performance.
However, the non-independent and identically distributed (non-IID) nature of data across different clients and the long-tail distributions, where some classes have significantly more samples than others, will reduce cache hit ratios and increase latency.
To address the aforementioned challenges, we propose an efficient inference framework, CoCa, which leverages a multi-client collaborative caching mechanism to accelerate edge inference.
On the client side, the model is pre-set with multiple cache layers to achieve a quick inference. During inference, the model performs sequential lookups at cache layers activated by the edge server. 
On the server side, CoCa uses a two-dimensional global cache to periodically aggregate information from clients, mitigating the effects of non-IID data.
For client cache allocation, CoCa first evaluates the importance of classes based on how frequently and recently their samples have been accessed.
CoCa then selects frequently recurring classes to address long-tail distribution challenges. Finally, CoCa dynamically activates cache layers to balance lookup overhead and accuracy.
Extensive experiments demonstrate that CoCa reduces inference latency by 23.0\% to 45.2\% on the VGG, ResNet and AST models with a slight loss of accuracy.

\end{abstract}
 

%

\section{Introduction}\label{sec:intro}
Deep neural networks (DNNs) have become increasingly prevalent in practical processing tasks on stream data, such as virtual/augmented reality (VR/AR) \cite{yi2020eagleeye, li2020object} and autonomous driving \cite{xu2021action, kumar2021omnidet, yu2020bdd100k}. 
These applications often require real-time processing of large volumes of data, posing significant challenges for traditional cloud-based solutions.
Traditionally, DNN-based inference tasks on stream data are managed via cloud servers, but this approach faces several significant challenges. 
First, offloading such tasks to servers entails substantial data transmission that will adversely affect the Quality of Service (QoS), such as response latency.
This is particularly problematic in latency-sensitive applications, such as autonomous driving, where delays will induce critical consequences. 
Second, there is a growing emphasis on privacy protection, exemplified by regulations such as the General Data Protection Regulation (GDPR) \cite{zaeem2020effect} in the European Union, which is one of the toughest privacy laws worldwide.
These considerations have catalyzed a shift toward edge computing \cite{cao2020overview, zeng2020coedge, shuvo2022efficient}, where inference processes migrate closer to data sources. 
This paradigm not only mitigates latency issues by reducing reliance on distant servers but also enhances data privacy through local data processing. 

The enhanced computing capabilities of edge devices have facilitated the proliferation of inference tasks on stream data, with driver assistance systems serving as a classic example.
These tasks always impose stringent Service-Level Objectives (SLOs).
For example, in driver assistance systems, achieving a response latency within 80 milliseconds, along with vehicle recognition accuracy of $\geq$ 95\% and traffic sign recognition accuracy of $\geq$ 92\% \cite{wali2019vision}, is essential to enhance the user experience.
To achieve these goals, a natural solution is to leverage the caching mechanism \cite{huynh2017deepmon,xu2018deepcache,li2021boosting}, which exploits the high temporal locality of video streams and the similarity between consecutive frames to accelerate inference.
Specifically, this approach stores the results of previous inferences in a cache, allowing quick access to avoid redundant computations for similar or identical frames in video streams.

Despite the potential of the caching mechanism, in order to boost inference performance, two primary challenges need to be addressed regarding data distribution.
(1) \textbf{Non-Independent and Identically Distribution (non-IID)}.
The data generated by clients often exhibit heterogeneity in distribution. For instance, videos captured by multiple cameras in a collaborative driving assistant task will vary greatly due to differences in camera positions, environments, and perspectives \cite{chiu2020semisupervised}. This heterogeneity makes it difficult to develop a unified caching mechanism that can effectively handle the diverse data features across different clients.
(2) \textbf{Long-Tail Distribution}.
The real-world data usually follow a long-tail distribution, where a small number of classes appear frequently, while a large number of classes appear infrequently. 
Specifically, in the context of automobile video analysis, normal driving behaviors are common and frequent, while abnormal behaviors are rare and diverse. This pattern results in a rich long-tail distribution \cite{zhang2021videolt, perrett2023use, ramesh2019sports}.
Traditional caching mechanisms, such as the Least Recently Used (LRU) strategy \cite{chrobak1999lru}, struggle to efficiently manage and retrieve data associated with these rare events, leading to suboptimal cache utilization and increased response latency.
These imbalance distribution characteristics pose significant challenges for edge inference acceleration. 

To address the aforementioned challenges, researchers have made significant efforts. 
LearnedCache \cite{balasubramanian2021accelerating} uses multiple exits and learned models to emulate caching operations, allowing early termination of inference upon prediction of cache hits, thus accelerating inference. 
LearnedCache attempts to adapt to the data distribution characteristics of clients through frequent retraining. 
However, the high computational overhead incurred by frequent retraining will degrade QoS, and even violate SLOs. 
Besides, there are often insufficient low-frequency class data for effective retraining under long-tail distributions.
To mitigate computational overhead, SMTM \cite{li2021boosting} employs low-dimensional representations (\ie, semantic information) of intermediate inference results as cache-matching entries.
This method frequently assesses the importance of each class based on the total frequency of occurrence and the most recent appearance time of their samples, effectively addressing the challenges of long-tail distributions.
However, the cache allocation strategy of SMTM often leads to increased lookup time and overall latency when multiple classes frequently appear in inference scenarios, which have also been validated in Section \ref{sec:evaluation}.

The above methods primarily focus on accelerating centralized inference in single-client environments. However, with the development of edge intelligence, collaboration among multiple clients is actively being explored to boost inference. 
FoggyCache \cite{guo2018foggycache} introduces cross-client approximate computation reuse with caching, emphasizing spatial-temporal correlation among client data. 
Initially, inference requests are processed locally; upon cache misses, queries are submitted to the server, and local caches are updated using the LRU replacement policy. 
FoggyCache addresses the non-IID issue by aggregating samples from different clients for server-side lookup entries, using cross-client data reuse to accelerate inference collectively. 
However, the LRU policy often fails under long-tail distribution, reducing the cache hit ratio and efficiency. 
In summary, existing approaches always fail to fully address the key challenges of non-IID and long-tail distributions.

We observe that in tasks such as smart city surveillance, data from spatially proximate devices often exhibit spatial similarity \cite{jiang2020federated}, which motivates us to leverage collaboration among multiple clients. 
To this end, we propose CoCa, which adopts a multi-client collaborative caching mechanism to exploit both temporal and spatial data similarities, overcoming the aforementioned challenges and accelerating inference.
Specifically, CoCa utilizes a client-server caching approach. The edge server maintains a two-dimensional global cache table, where the columns correspond to the preset cache layers of the model, and the rows correspond to the entries of classes at different layers. 
CoCa periodically aggregates information from clients to update the global cache at the edge server, mitigating the effects of non-IID data and capturing contextual feature changes in the client. 
The server then allocates personalized caches to each client by extracting sub-tables from the global cache, corresponding to entries of frequently recurring classes (termed hot-spot classes) at specific cache layers.
To address the long-tail distribution challenge, we assess hot-spot classes based on a combination of the overall class frequency and recent occurrence patterns.
Upon receiving the cache, clients perform model inference with caching.
During inference, the model sequentially performs cache lookups at each activated cache layer, calculating cumulative scores for each class and checking for cache hits. If a cache hit occurs, the inference process is terminated early, and the cached result is returned.

CoCa offers significant advantages over existing cache-based methods. 
Specifically, the multi-client collaborative caching mechanism in CoCa effectively addresses non-IID and long-tail distribution challenges through global cache updates and adaptive client-side cache allocation.
However, two main challenges persist in achieving efficient inference with CoCa.
Firstly, the number and location of active cache layers in the model significantly affect the cache lookup cost and effectiveness.
Too few cache layers will reduce cache hit ratio and acceleration opportunities, while too many layers will increase average inference latency (see Section \ref{sec:motivation 2}). Thus, \textit{activating the appropriate cache layers in the model (\ie, columns in the local cache) is crucial for balancing accuracy performance and lookup overhead}.
Secondly, we observe that the number of hot-spot classes varies a lot under different environments. In scenarios where a small number of classes frequently appear, using a cache with a fixed large number of classes may dilute the benefits of cache hits due to time wasted on lookups. Conversely, when the number of hot-spot classes exceeds the number of classes in the cache, it will lead to a high cache miss ratio or even erroneous hits. Therefore, another challenge is \textit{how to dynamically adjust the number of hot-spot classes in the cache entries (\ie, rows in the local cache) to improve inference accuracy and reduce inference latency.}
The main contributions are summarized as follows:

\begin{itemize}
    \item We propose an efficient framework CoCa, which leverages a multi-client collaborative caching mechanism, to enhance edge inference.
    CoCa performs global cache updates on the server and dynamically allocates cache entries for each client to handle non-IID and long-tail distributions. 
    
    \item To address the challenges in CoCa, we propose an efficient heuristic cache allocation algorithm to dynamically determine the appropriate cache layers and the number of hot-spot classes in the cache based on the data distribution and the recent local class occurrence patterns of each client. 

    \item The extensive experimental results demonstrate that CoCa significantly reduces average inference latency, achieving reductions of 23.01\% to 45.19\% the VGG, ResNet and AST models, with a slight ($<$ 3\%) accuracy loss. 
    Furthermore, CoCa achieves latency reductions of 16.67\% to 21.26\% on ResNet101 compared to other baselines.
    
\end{itemize}

\section{Background and Related Work}\label{sec:motivation}
\subsubsection{Edge Inference}
Traditional machine learning (ML) inference has primarily been offloaded to centralized servers, taking advantage of their substantial processing resources \cite{alfakih2020task, chen2022context, zhang2021elf}. 
However, contemporary concerns regarding data privacy and the increasing demand for minimal inference latency have catalyzed a new trend. 
Edge devices (or clients), such as mobile phones, are emerging as feasible platforms for executing inference tasks, thus eliminating the need to offload these computations to the server. 
In the realm of edge inference, tasks such as virtual/augmented reality and autonomous driving, with image classification as a predominant domain, have garnered significant focus.
The existing research has increasingly concentrated on optimizing model executions to enhance inference speed on edge devices, exploring a variety of optimization strategies \cite{guo2018foggycache, cheng2017quantized, chen2021accelerating}. 
A prevalent strategy involves reducing the computational workload per inference through model compression, model partitioning and edge-cloud collaborative inference, which accelerates the inference process. 
For instance, Cheng \etal \cite{cheng2017quantized} demonstrate that weight quantization can significantly mitigate the storage and computational burdens of models, thus reducing inference latency. 
Chen \etal \cite{chen2021accelerating} propose a scheme to dynamically partition DNN into two or three parts and distribute them at the client and server, achieving the lowest delay with the change of request rate. 
These methods typically focus on point-to-point collaboration between individual clients and servers, overlooking collaboration optimization across multiple clients. 
\subsubsection{Cache-based Inference}
Incorporating caching mechanisms in model inference is a widely adopted technique to reduce latency and enhance long-term throughput.
This approach stems from a crucial observation regarding the temporal locality often present in tasks with stream data \cite{huynh2017deepmon, xu2018deepcache}. 
Such tasks frequently exhibit a pronounced temporal coherence among consecutive frames within a video stream, where successive frames often manifest high similarity and correlation in their content and resultant classifications. 
To capitalize on this temporal locality inherent in continuous vision tasks, a straightforward approach involves the incorporation of caching. 
A common caching strategy typically involves inserting caches at the entry or intermediate layers of the model.
When inference reaches the cache, it searches for the most similar prior entries, using them for result determination through methods like voting upon a cache hit.
Extensive scholarly research on caching mechanisms for DNN models has produced varied proposals.
DeepMon \cite{huynh2017deepmon} pioneers in using cache by leveraging the high temporal locality of stream data, reusing partial block intermediate results from previous frames through caching, thus reducing computational load and accelerating inference. 
DeepCache \cite{xu2018deepcache} advances based on DeepMon by incorporating inter-frame motion characteristics, enabling the reuse of non-corresponding position blocks. 
SMTM \cite{li2021boosting} introduces an innovative caching mechanism, using the centroid of intermediate result vectors from same-class samples as cache entries and employing cosine similarity for matching during inference. 

\subsubsection{Class-based Semantic Caching}
To reduce storage consumption and complexity in caching, we primarily employ a class-based semantic caching mechanism \cite{li2021boosting} in this paper.
During inference, the model is partitioned into multiple blocks based on preset cache locations, with cache layers set between these blocks.
The cache layers are populated with low-dimensional feature representations of hot-spot classes, resulting in a relatively small overall cache size. 
At any cache layer $j$, each cache entry corresponds to the semantic vector center of a class in the classification task, and the model extracts a one-dimensional semantic vector through global average pooling. 
Then, we calculate the cosine similarity between the semantic vector and the cache entry of hot-spot class $i$, denoted as ${C}_{i,j}$.
To enhance discriminative robustness, the cross-layer accumulated cosine similarity ${A}_{i,j}$ of class $i$ in layer $j$ is used, as follows:
\begin{equation}
    {A}_{i,j} = {C}_{i,j} + \alpha \cdot {A}_{i,j-1}
\end{equation}
Note that, a decay coefficient $\alpha$ (set to 0.5 by default) is employed to diminish the weight of previous cache layers. 
Assuming classes $a$ and $b$ have the first and second largest cumulative cosine similarity in cache layer $j$, the cross-layer accumulated cosine similarities of classes $a$ and $b$ are denoted as ${A}_{a,j}$ and ${A}_{b,j}$. Then the discriminative score ${D}_j$ of the current layer $j$ is calculated as:
\begin{equation}
    {D}_j = \frac{ {A}_{a,j} - {A}_{b,j} }{ {A}_{b,j} } 
\end{equation}
The cache hits if the discriminative score ${D}_j$ exceeds the predefined threshold $\bm{\Theta}$. In this case, class $a$ is outputted as the classification result for this iteration and the inference process terminates. If the cache misses, the inference process continues.


\section{Motivations for Framework Design}\label{sec:motivation 2}




In this section, we first analyze the impact of the cache size on inference latency. 
Then, we explore the influence of the number and locations of cache layers and the number of hot-spot classes on the inference performance, such as hit ratio and accuracy. Finally, we compare the effectiveness of the cache with and without global updates in multi-client settings.


\textit{\textbf{1) Increasing cache size initially reduces latency, but excessively large caches lead to increased latency.}}
In semantic cache inference, caches can be set after each convolutional layer of the model, allowing for multiple cache layers within the model. For instance, in a ResNet101, up to 34 cache layers can be inserted. These caches have a relatively small size; for the UCF101 task, the total cache size is approximately 3.2 MB, which is 1.95\% of the model size \cite{li2021boosting}.
Considering the small cache size and usage, the impact of transmission delay on the overall inference latency is negligible. 
Despite the small memory footprint, the computational overhead for cache lookups is significant. 
Based on our experiments, the total lookup latency for all 34 cache layers amounts to 56.22\% of the inference latency without caching.

\begin{figure}[t]
  \centering
  \subfigure[Performance with different cache sizes.]{\includegraphics[width=1.66in]{fig/2.2.pdf}\label{fig:1a}}
  \hspace{0.5mm}
  \subfigure[Performance of varied cache layers.]{\includegraphics[width=1.7in]{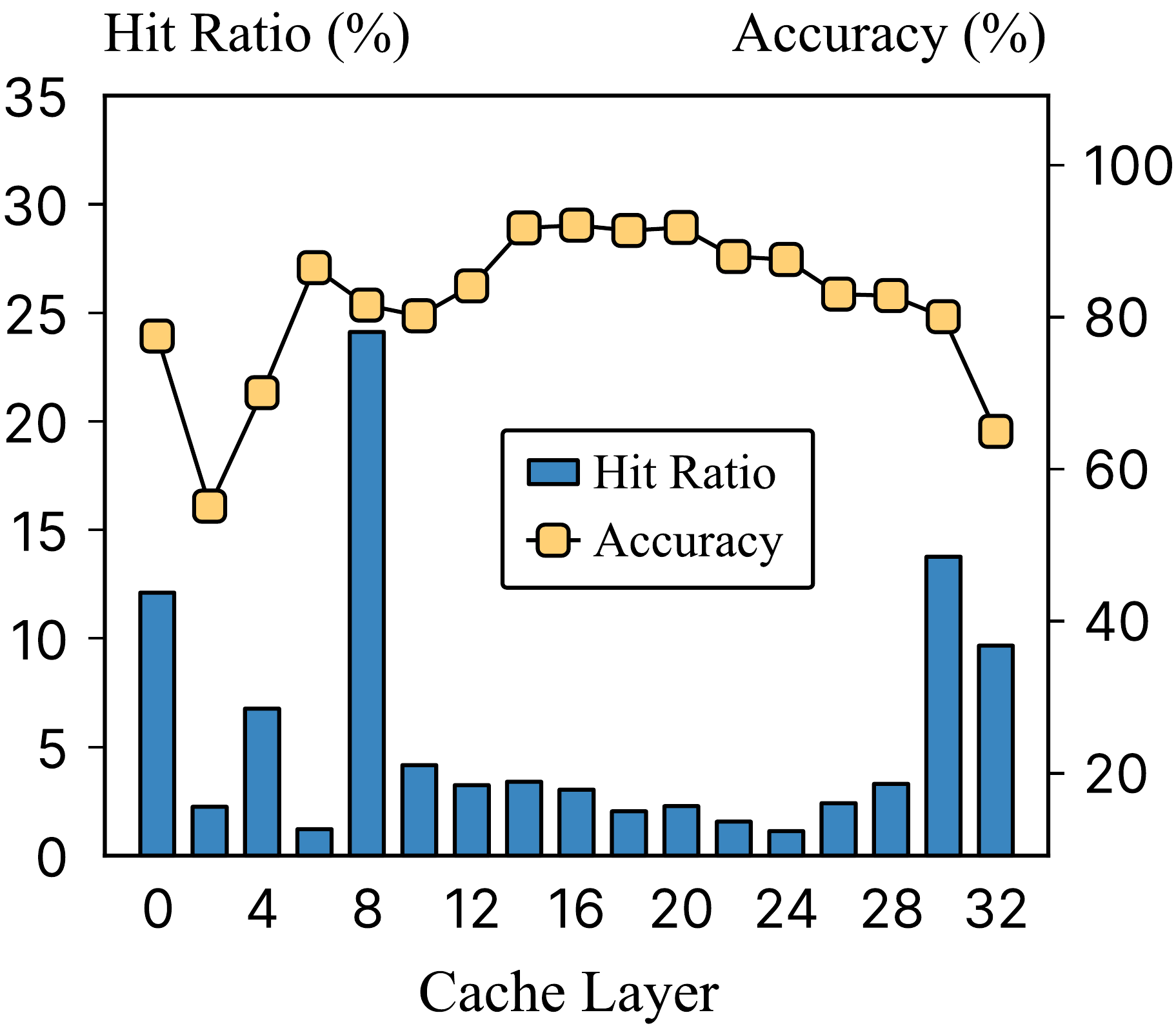}}
  \caption{The test result of ResNet101 on a subset of 50 classes from UCF101. The left plot shows the latency and accuracy performance of settings with different cache sizes. The right plot shows the hit ratio and accuracy performance of different cache layers.}
  \label{fig:fig_1}
\end{figure}

In order to investigate the impact of cache size on inference performance, we adopt ResNet101 on a subset of 50 classes from the UCF101 dataset with different cache sizes.
The cache size is determined by the number of cache layers and the number of hot-spot classes in the cache. 
To mitigate the influence of the cache entry selection algorithm, we set the number of hot-spot classes to the total number of classes. 
We control the cache size by adjusting the number of cache layers and activating them at regular intervals across the 34 layers. 
%
We set the maximum cache size (\ie, 100\% in Fig. \ref{fig:1a}), which is approximately 3.2 MB when all 34 layers are activated, as our baseline.
Then we use the percentage of this baseline as the horizontal axis for an easy understanding of the relative relationships among different settings.
The average inference latency and accuracy are shown in Fig. \ref{fig:fig_1}(a). 
As the cache size varies, inference accuracy remains stable, with the accuracy loss consistently below 2\%. 
Initially, as the cache size increases, the average inference latency rapidly decreases, reaching its lowest observed point when the activated cache size is 10\% of the total cache size. This configuration reduces the average latency by 28\% compared to the setting without caching. As the cache size continues to increase, the latency gradually increases slightly.
These results indicate that using a cache can effectively reduce the average inference latency while incurring a low accuracy loss. However, a larger cache size does not necessarily improve performance due to increased lookup time. Therefore, it is crucial to judiciously control the cache size to balance lookup overhead and cache hit benefits, achieving optimal performance.

\textit{\textbf{2) Different cache layers and the number of hot-spot classes exhibit varying lookup time, hit ratio, and accuracy.}}
Based on previous experiments, we conclude that constraining cache size within a certain threshold is essential for effective latency reduction.
We can adjust the cache size to meet the threshold from two aspects: the number of cache layers and the number of hot-spot classes in the cache.
Model inference with multiple cache layers can increase the overall cache hit ratio, but it results in longer cache lookup time. 
Similarly, increasing the number of hot-spot classes will enhance the likelihood of cache hits, but prolong lookup time.
To investigate the effect of these two factors in detail, we conduct experiments using the previously described model and dataset.

First, we measure the hit ratio and the accuracy following a cache hit at each individual cache layer when all cache layers are activated.
As shown in Fig. \ref{fig:fig_1}(b), the hit ratio is high in shallow and deep cache layers but low in middle layers. This pattern emerges because easily inferrable samples are hit early. In deeper cache layers, intermediate features become more distinctive, increasing the hit ratio. Middle cache layers naturally have lower hit ratios. 
The accuracy of cache hits is lower in both shallow and deep cache layers, compared to intermediate layers. This can be attributed to insufficient feature discriminability in shallow layers, while deep layers hit mainly difficult samples, resulting in decreased accuracy.
Shallow cache layers offer higher latency reduction benefits, while deep layers provide better accuracy. 
Therefore, these observations underscore the necessity for careful selection of cache layers to optimize the trade-off between latency reduction and accuracy.

\begin{table}[t]
\centering
\setlength{\abovecaptionskip}{10pt}%
\setlength{\belowcaptionskip}{0pt}%
\footnotesize
\caption{The average inference latency (Lat.) and accuracy (Acc.) of ResNet101 with different numbers of hot-spot classes in the cache on a subset of 50 classes from UCF101 and ImageNet-100 (0 means without cache).}
\label{table2}
\begin{tabular}{ c c c c c}
\hline
\rule{0pt}{8pt}
           Number of & \multicolumn{2}{c}{UCF101} &  \multicolumn{2}{c}{ImageNet-100} \\
\cline{2-5}
\rule{0pt}{8pt}
hot-spot classes  & Lat.(ms) & Acc.(\%) & Lat.(ms) & Acc.(\%) \\
\hline
\rule{0pt}{8pt}
0   & 40.58 & 80.56  & 40.33  &  82.07   \\
\hline
\rule{0pt}{8pt}
10   & 23.56 & 47.51  & 25.98  & 35.40    \\
\hline
\rule{0pt}{8pt}
30   & 29.23 & 61.45  & 28.70  & 62.81    \\
\hline
\rule{0pt}{8pt}
\textbf{50}   & \textbf{30.53} & \textbf{80.08}  & \textbf{29.08}  & \textbf{81.46}    \\
\hline
\rule{0pt}{8pt}
70   & 33.36 & 80.12  & 33.04  & 81.47    \\
\hline
\rule{0pt}{8pt}
90   & 36.04 & 80.07  & 35.45  & 81.21    \\
\hline
\end{tabular}
\end{table}

Second, we test the effect of the number of hot-spot classes in the cache on inference performance.
Based on the previous experiment setting, we add ImageNet-100 to enrich our results.
From the test results shown in Table \ref{table2}, as the number of hot-spot classes increases, both latency and accuracy rise initially. When the number of hot-spot classes exceeds 50, accuracy stabilizes, with a loss of less than 1\% compared to the result of setting without caching. However, latency begins to increase, diminishing the latency reduction benefits of caching.
This trend reflects the trade-off between cache hit ratio and lookup time. Initially, increasing hot-spot classes improves the hit ratio and accuracy but also increases lookup time meanwhile. The optimal balance occurs around 50 classes, after which further expansion yields diminishing returns on hit ratio and accuracy while continuing to increase latency.
In summary, it is pivotal to appropriately select cache layers and the number of hot-spot classes for optimizing cache efficiency and ensuring accurate inference results.

\textit{\textbf{3) Multi-client collaborative global updates exhibit a positive impact on cache accuracy improvement.}}
In the semantic caching mechanism, the cache entry (\ie, low-dimensional semantic center) of each class at each cache layer is crucial for cache lookups. 
Cache hits and inference outcomes depend on the similarity between the intermediate semantic vector of input and cache entries. The accuracy of cache hits depends on the representational power of cache entries.
Moreover, client data often exhibit non-IID characteristics, such as varying image data from traffic intersections and pet shop cameras. In some tasks, data from multiple clients have spatial similarities. 
Inspired by federated learning \cite{li2021efficient, qu2021federated}, we periodically aggregate semantic vectors from different clients into a global cache table to leverage these diverse datasets. 
We collect inference samples from clients and use their computed semantic vectors from each cache layer to update the global cache table. 
The global updates are periodically performed via weighted summation of new semantic vectors (derived from various clients) and corresponding entries of the same class and layer in the global cache table.
This process incorporates diverse semantic information from various clients into the global cache, enhancing its representational power.

\begin{figure}[t]
    \centering
    \includegraphics[width=3in]{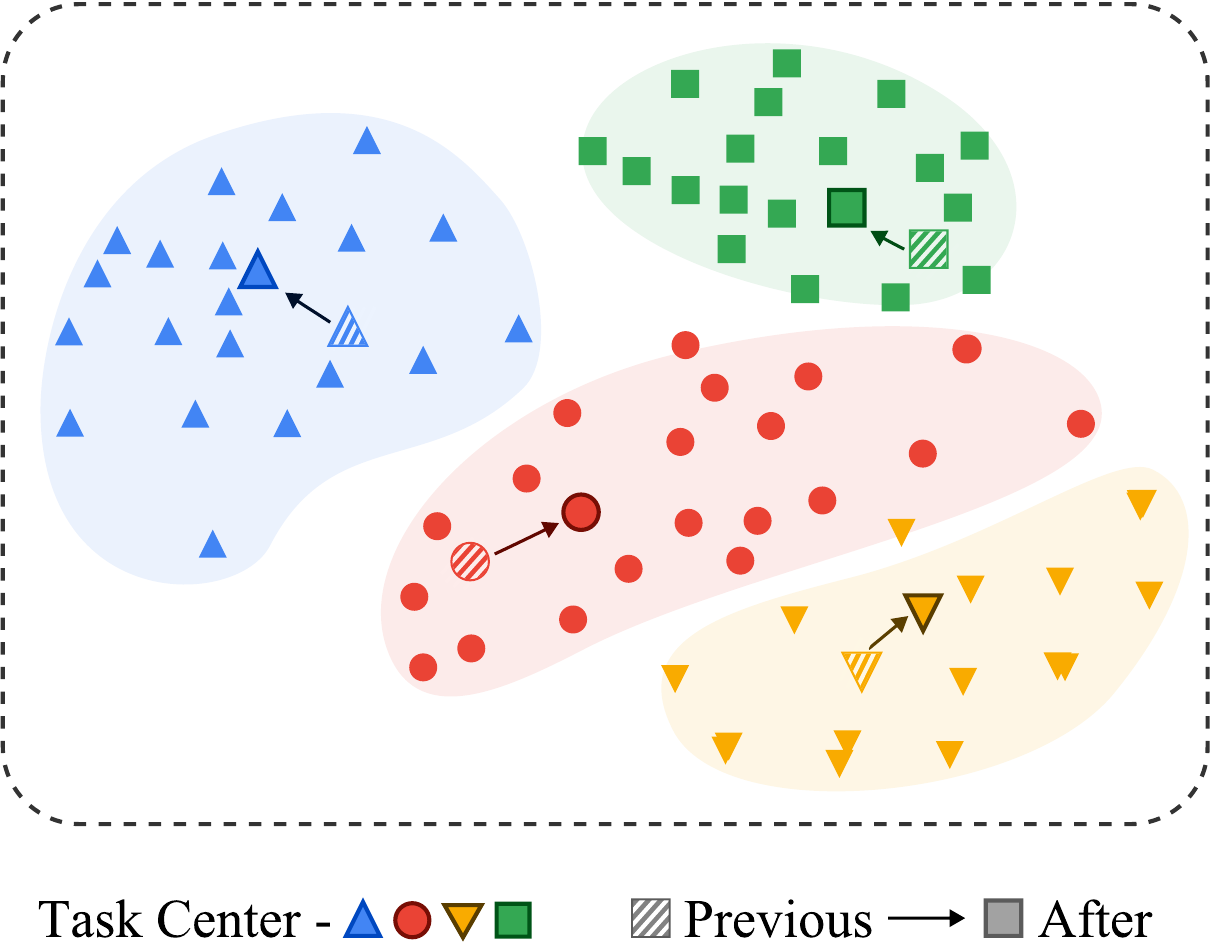}
    \caption{The t-SNE visualization of cosine similarity clustering for the semantic vectors of samples from 4 classes of the UCF101 dataset, cached in the 18-th layer of all 34 cache layers in ResNet101. The larger points represent the cached semantic centers for each class. "Previous" and "After" show the cached entries before and after employing the global updates strategy, respectively.}
    \label{fig:enter-label}
\end{figure}

To investigate the effect of global updates on the representational power of cache entries and inference accuracy, we conduct experiments using 10 clients.
Our experiments use ResNet101 on a subset of 20 classes from UCF101, tracing the performance of the cache under settings with and without global updates. 
The server generates the initial cache using the global shared dataset. Under the setting with global updates, we update the cache entries based on inference samples from the clients.
An equal number of samples of each class are drawn from a client for testing. 
We compare the clustering effects of the cosine similarities between the cache entries and the semantic vectors of samples (from the 18-th layer of all 34 cache layers in ResNet101), with and without global updates. Fig. \ref{fig:enter-label} illustrates the t-SNE 2D visualization of their relative distances. We plot the changing trends of the semantic centers for four classes under global updates as an example.  
It can be observed that with global updates, the semantic centers of each class become more closely aligned with their corresponding class sample centers, resulting in improved clustering effects. 
The inference accuracy with global updates is also higher than the result without global updates, which is validated in Section \ref{ablation}. 
These results indicate that the global cache helps align the semantic centers of the cache more closely with the current data feature, positively impacting the hit ratio and accuracy of cache lookups during inference.



\section{Proposed Framework }\label{sec:framework}

\subsection{System Overview}
Inspired by the above observations, we introduce CoCa, an efficient inference framework with a multi-client collaborative caching mechanism. 
The workflow of CoCa for each client is illustrated in Fig. \ref{fig:frame}.
The server first deploys pre-trained models to each client. Then, it iteratively performs the following steps round-by-round.
%
%
At the beginning of each round, each client uploads its status information, including the current data class distribution and timestamp vectors, to the server. The server then allocates the appropriate cache to each client based on the global commonality information and the client's status information.
Subsequently, the client loads the received cache and performs local inference on a set of frames. During this process, it collects client status information and semantic vectors of inference samples.
At the end of the round, each client uploads the information collected during the current round to the server for global updates. These updates include refreshing the global cache table and class frequency information. Following this, the system prepares for the next round.


\begin{figure}[t]
  \centering
  \includegraphics[width=0.8\linewidth]{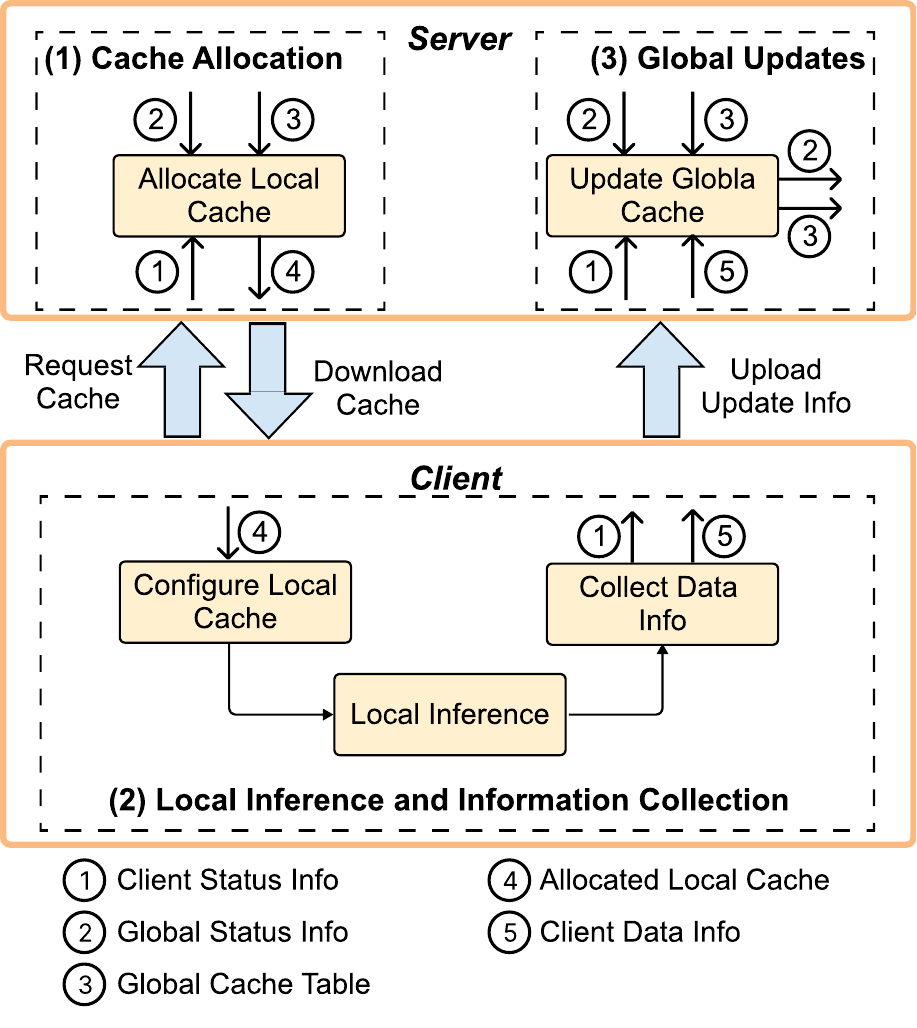}
  \caption{Overview of CoCa.} 
  \label{fig:overview}
\end{figure}

\subsection{Cache Allocation}

For cache allocation request of client $k$, the server allocates suitable cache entries from the global cache, based on the client status information collected (such as the local timestamp vector $\bm{\tau}^k$) and the global status information maintained on the server. Details of the cache allocation algorithm can be found in Section \ref{sec:algorithm}.
%
In simple terms, upon receiving the client's cache request and status information, the server allocates the appropriate cache through a two-step process. First, based on the client and global data status information, the server selects the most likely to appear classes to form a hot-spot classes set. Second, the server chooses cache layers in order of expected benefits (such as expected latency reductions) from high to low, and fills these layers with corresponding cache entries from the hot-spot classes set until the allocated cache size approaches the client's cache size threshold. Accordingly, the server successfully completes the cache allocation for the client.

\subsection{Client Information Collection}
In the second step of a single round, the client performs local collection of client status information and gathers semantic vectors of inference samples during local inference.
The client initializes two one-dimensional vectors to track the recent occurrences of data classes: a timestamp vector $\bm{\tau}$ and a local class frequency vector $\bm{\varphi}$.
Specifically,
we use ${\tau}_{i}$ (the $i$-th element of $\bm{\tau}$) to represent the number of inference processes since the last appearance of a sample of class $i$. After each inference process, the timestamp of the current sample class is reset to zero, while the timestamps of other classes are incremented by one.
We use ${\varphi}_{i}$ (the $i$-th element of $\bm{\varphi}$) to record the number of times that samples of class $i$ have appeared within the current round.
These vectors enable the client to maintain a dynamic record of recent class occurrences.
CoCa operates in rounds, processing $F$ frames in each round. Based on our tests, we set $F = 300$ to achieve stable periodic cache updates and uploads of client information.
During local inference, the client maintains status information $\bm{\tau}, \bm{\varphi}$ and collects the cache update table $\mathbf{U}$ for global updates.
The cache update table has the same dimensions as the global cache table on the server. 
The client determines whether to collect information from the current inference process to populate $\mathbf{U}$ based on the following strategy.
Specifically, two types of inference samples will be chosen:
\begin{enumerate}
    \item Samples where the cache hits and the discriminative score ${D}_j$ (hits at the cache layer $j$) exceeds the threshold ${\Gamma}$. 
    These samples exhibit similar features to the current cache and can be used to reinforce it, thereby capturing the gradual evolution of class semantics over time. The collected semantic vectors are limited to the point of the cache hit.

    \item Samples where the cache misses and their normalized probability vectors $\mathbf{prob}$ of classes satisfy the following condition: the difference between the two largest probabilities, \ie, ${prob}_1 - {prob}_2$, is greater than the threshold ${\Delta}$. These samples display high confidence in the classification results but may not be adequately captured by the current cache, serving as a supplement to the global cache.
\end{enumerate}
These two types of samples represent the reinforcement and expansion of the current cache entries, respectively. The balance between reinforcement and expansion can be adjusted by tuning the two thresholds ${\Gamma}$ and ${\Delta}$.

Now, we will describe the method for collecting semantic vectors.
For the selected inference sample of class $i$, its semantic vector at layer $j$ is denoted as $\mathbf{V}_{i,j}$, and the corresponding cache entry in the cache update table is denoted as $\mathbf{U}_{i,j}$. 
The entry $\mathbf{U}_{i,j}$ is first updated as follows:
\begin{equation}
    \mathbf{U}_{i,j} = \mathbf{V}_{i,j} + {\beta} \cdot \mathbf{U}_{i,j}
    \label{eq:update U}
\end{equation}
where ${\beta}$ is the decay coefficient (set to 0.95 by default), which attenuates the influence of older inference samples.
Then, $\mathbf{U}_{i,j}$ is normalized to have a unit L2 norm.
Upon completing local inferences and collecting update information, the client uploads the cache update table for global updates. The client continues to the next round, passing the task back to the server side.

\subsection{ Global Cache Updates }
Global updates primarily involve updating the global cache table and class frequency information. For global cache updates, both the global cache table and the cache update table uploaded from the clients have the same two-dimensional logical structure. 
In CoCa, the global cache table is updated by element-wise weighted summation of the corresponding entries from the local cache update table and the global cache table, with weights based on local and global class frequencies.
For client $k$, we denote the entry in the local cache update table for class $i$ and cache layer $j$ as $\mathbf{U}^k_{i,j}$, and the entry in the global cache table as $\mathbf{E}_{i,j}$.
Corresponding to the local class frequency vector $\bm{\varphi}^k$, the global class frequency vector is denoted as $\bm{\Phi}$.
For the update information from client $k$, the global cache entry $\mathbf{E}_{i,j}$ is updated as follows:
\begin{equation}
    \mathbf{E}_{i,j} = {\gamma} \cdot \frac{ {\Phi}_{i} }{ {\Phi}_{i} + {\varphi}^k_{i} } \cdot \mathbf{E}_{i,j} + \frac{ {\varphi}^k_{i} }{ {\Phi}_{i} + {\varphi}^k_{i} } \cdot \mathbf{U}^k_{i,j}
    \label{eq:update E}
\end{equation}
where ${\Phi}_{i}$ and ${\varphi}^k_{i}$ represent the global total frequency and local frequency, respectively, for the samples of class $i$. 
The parameter ${\gamma}$ is the decay coefficient, and is set to 0.99 for subsequent experiments. Thus, we evaluate the importance of updating entries using their frequency and perform a weighted average. 
After updates, $\mathbf{E}_{i,j}$ is normalized to have a unit L2 norm.
The global class frequency vector $\bm{\Phi}$ is updated by adding the local class frequency vector $\bm{\varphi}^k$ of client $k$. Mathematically, the update for each class $i$ is formulated as:
\begin{equation}
    {\Phi}_i = {\Phi}_i + {\varphi}^k_i
    \label{eq:update phi}
\end{equation}
The procedure outlined above describes how the server implements the updates of the global cache and state information.

\begin{figure}[t]
  \centering
  \includegraphics[width=\linewidth]{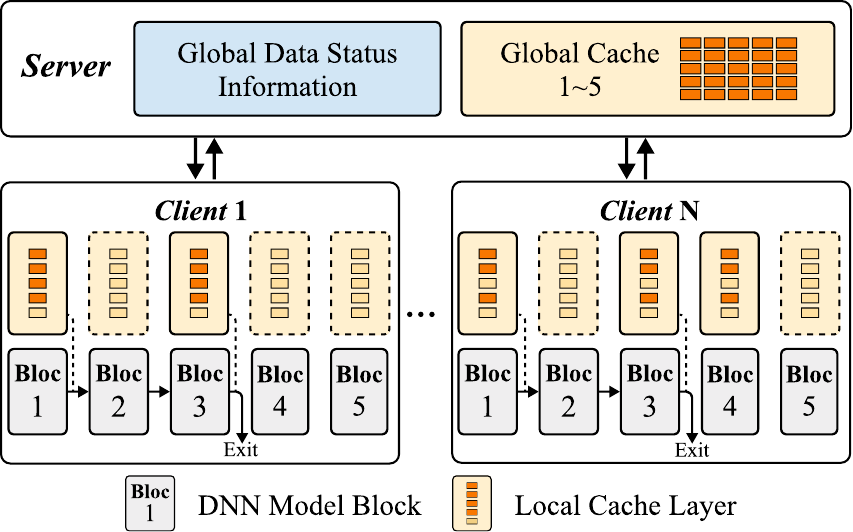}
  \caption{An illustrative example of CoCa at round $T$. The server maintains a 5×5 global cache. Rows correspond to different classes, while columns correspond to different cache layers. Dashed outlines indicate empty cache layers, and dark blocks represent semantic cache entries.}
  \label{fig:frame}
\end{figure}

\subsection{An Illustrative Example of CoCa}
To illustrate our proposed framework CoCa, we take the example depicted in Fig. \ref{fig:frame}. 
The server maintains a $5 \times 5$ two-dimensional global cache table. The five rows represent the five classification classes of the task, and the five columns correspond to the five pre-set cache exit layers in the model.
Consider a scenario with a cache size threshold of ten entries, 
The recent data stream in Client 1 is heavily concentrated on four primary classes, which have more recent timestamps and higher frequencies compared to the infrequent appearance of other classes. 
Upon receiving the state information and cache request of Client 1, the server allocates four entries per cache layer. 
Then, the server uses a greedy algorithm to sequentially select cache layers that maximize contribution. 
As a result, the server may provide Client $1$ with the first and third cache layers, each populated with four entries corresponding to the frequently accessed classes.
When the number of frequently accessed classes for Client $N$ is three, then the server adapts by selecting three cache layers (such as the first, third, and fourth layers), each containing three entries.
This approach dynamically accommodates the changing data patterns of clients. As the data status of clients continuously evolve, the server adaptively allocates appropriate caches in each round.
In CoCa, each client communicates solely with the server, eliminating the need for direct synchronization of cache updates and allocations between clients. 
Information sharing across clients is facilitated through the server’s global cache update mechanism.
 



\section{Problem Formulation and Algorithm Design}\label{sec:algorithm}
\subsection{Problem Formulation}

In this section, we define the cache entries allocation optimization problem in our proposed framework CoCa. The goal is to minimize average inference latency under constraints of cache size and inference accuracy loss on the client side through optimal allocation of cache entries.
Assuming there are ${N}$ clients, we need to allocate cache entries for each client from the global cache table. The global cache table consists of $I$ rows and $L$ columns, where $I$ represents the total number of distinct classes in classification tasks, and $L$ represents the maximum number of cache layers.
Let ${x}_{i,j}^k$ denote the allocation status of the cache entry in row $i$ and column $j$ for client $k$, where ${x}_{i,j}^k \in \{0, 1\}$ for $0 < i \leq I$, $0 < j \leq L$, and $0 < k \leq {N}$.
The size of the local cache is directly affected by the allocation status ${x}_{i,j}^k$. Specifically, ${x}_{i,j}^k = 1$ indicates that the corresponding cache entry is allocated, while ${x}_{i,j}^k = 0$ means it is not. Therefore, the total size of the local cache is calculated by summing the sizes of all allocated cache entries.
Considering the varying sizes of the entries in different cache layers, we formulate the local cache size for client $k$ as follows:
\begin{equation}
    M^k = \sum_{0 < i \leq I, 0 < j \leq L} {x}_{i,j}^k \cdot m_{i,j}
\end{equation}
where $m_{i,j}$ denotes the size of the entry in row $i$ and column $j$ of the cache table, \ie, the cache entry of class $i$ at layer $j$.

During inference, when the cache hits, the inference result corresponds to the class of the cache entry hit. Consequently, increasing the number of stored entries in the cache enhances the likelihood of the correct class being present. This, in turn, reduces the inference errors arising from cache hits where the correct class entries are absent, thereby decreasing the loss in inference accuracy.
On one hand, the accuracy loss is influenced by the number of entries per cache layer, which is related to the elements of $\mathbf{X}^k$. The inference accuracy loss of client $k$ is denoted by $\mathcal{G}^k(\mathbf{X}^{k}, {\Theta}^k)$, where $\mathbf{X}^{k}$ represents the set of all ${x}_{i,j}^k$, and ${\Theta}^k$ denotes the cache hit threshold for client $k$, respectively. 
Given the complexity of deriving an exact formula for $\mathcal{G}^k(\mathbf{X}^{k}, {\Theta}^k)$, we estimate and characterize this relationship using empirical data obtained from server-side testing on a shared dataset.
On the other hand, ${x}_{i,j}^k$ exhibits a positive correlation with cache size, subsequently resulting in a prolonged cache look-up time.
This is intuitive: larger cache allocations result in more cache layers and entries per layer to be searched, leading to higher computational costs for cache lookup.
The $L$ cache layers divide the model into $L+1$ model blocks. Let ${\Lambda}_{j}$ and $\mathcal{C}_{j}(\mathbf{X}^{k})$ denote the computation time of the model block $j$ and the lookup time of the cache layer $j$, respectively. As there is no cache layer $L+1$, we set $\mathcal{C}_{L+1}(\mathbf{X}^{k})=0$. We define $\mathcal{P}_{j}(\mathbf{X}^{k}, {\Theta}^k)$ as the probability of a cache hit before (not including) the cache layer $j$. The relationship between the hit ratio and the cache can be represented by an empirical relation tested on a shared dataset by the server.
Thus, the expected inference time for client $k$ can be formulated as:
\begin{equation}
    \mathcal{T}^k(\mathbf{W}^k) = \sum_{j=1}^{{L+1}} (1-\mathcal{P}_{j}(\mathbf{X}^{k}, {\Theta}^k)) \cdot ({\Lambda}_{j} + \mathcal{C}_{j}(\mathbf{X}^{k}))
\end{equation}

When allocating the local caches, CoCa allocates the appropriate cache entries for each client to reduce the average inference delay for all clients. Assuming each client $k$ has ${n}_k$ samples, the global average inference latency $\mathcal{T}(\mathbf{W})$ can be defined as:
\begin{equation}
    \mathcal{T}(\mathbf{W}) = \frac{1}{\sum_{k=1}^{{N}} {n}_k} \cdot \sum_{k=1}^{{N}} {n}_k \cdot {\mathcal{T}}^{k}(\mathbf{W}^k) 
\end{equation}
where $\mathbf{W}$ is the set of parameters of all clients, defined as:
$\mathbf{W} = \{\mathbf{W}^1, \mathbf{W}^2, ..., \mathbf{W}^N\}$.
Here, $\mathbf{W}^k$ represents the parameter set for client $k$.
We denote the number of samples and the average inference latency at client $k$ as ${n}_k$ and ${\mathcal{T}}^{k}(\mathbf{W}^k)$, respectively.
Accordingly, when sample numbers ${n}_k$ of each client are consistent, we can simplify the optimization problem as follows:
$$ \min_{\{\mathbf{W}\}}  \sum_{k=1}^{N} {\mathcal{T}}^{k}(\mathbf{W}^k) $$ 
\begin{equation}
    s.t. \left\{
    \begin{aligned}
        & \sum_{i \in \mathbf{I}, j \in \mathbf{L}} {x}_{i,j}^k \cdot m_{i,j} \leq {\Pi}^k   \qquad & \forall k  \\
        & \mathcal{G}^k(\mathbf{X}^{k}, {\Theta}^k) \leq {\Omega}^k  \qquad & \forall k  \\
        & {x}_{i,j}^k \in \{0, 1\}     \qquad & \forall i,j,k  \\
    \end{aligned}
    \right.
\end{equation}

The object of the optimization problem is to minimize the average inference latency of all clients. The first set of inequalities expresses the cache size constraint for every client, where ${\Pi}^k$ is the cache size threshold prescribed for the client $k$. The second set of inequalities expresses the inference accuracy loss constraints for every client, where ${\Omega}^k$ is the accuracy loss threshold determined for the client $k$. 
The third set of constraints defines ${x}{i,j}^k$ as binary variables, where ${x}_{i,j}^k \in \{0, 1\}$ for all $i, j, k$.
${x}_{i,j}^k = 1$ means that the cache entry in row $i$ column $j$ of the global cache table is allocated to the client $k$, and 0 means the opposite. 

\subsection{Algorithm Design}

%
The optimization problem defined above is a combinatorial optimization problem that can be hard to solve directly  \cite{papadimitriou2013combinatorial}. The complexity of this problem grows exponentially with the number of clients, cache entries, and layers, rendering exact solutions computationally intractable for practical instances.
Given this inherent complexity, we propose a heuristic \textit{Adaptive Cache Allocation (ACA)} Algorithm to efficiently obtain a suboptimal cache allocation scheme. 
We will detail ACA in the following sections.

\subsubsection{\textbf{Algorithm Preliminary}} 

The model architecture incorporates $L$ cache layers. When making cache entry allocation decisions, the server utilizes the global class frequency vector $\bm{\Phi}$ and the local class timestamp vector $\bm{\tau}^k$ uploaded from the client $k$. Here, $\bm{\Phi}_i$ and $\tau_i^k$ represent the values of the class $i$ in each vector, respectively. The global class frequency vector $\bm{\Phi}$ records the occurrence frequency of each class in the inference samples, reflecting the differences in the appearance frequencies of each class. The local class timestamp vector $\bm{\tau}^k$ records the number of consecutive inference samples in which each class has not appeared on the client side currently, indicating the recency of each class. After each inference, $\tau_i^k$ is reset to zero when the sample of class $i$ appears, otherwise incremented by one. 
The parameter $F$ represents the cycle of client cache's update. For every $F$ inferences, clients initiate cache entry allocation requests to the server anew. The client also uploads a cache size constraint threshold ${\Pi}^k$ to guide the allocation of cache entries.

The server also receives cache hit ratio vector $\mathbf{R}$ and saved inference time vector $\bm{\Upsilon}$ uploaded by the clients. Both vectors have a length of ${L+1}$. The cache hit ratio vector $\mathbf{R}$ records the hit ratio of each cache layer and the saved inference time vector $\bm{\Upsilon}$ records the saved inference time if a cache hit occurs at each cache layer (considering model computation time only). Combining the contents of these two vectors helps measure the importance of different cache layers, thus assisting in retaining more crucial cache entries during cache entry allocation.

\subsubsection{\textbf{Algorithm Description}}

\begin{algorithm}[t]
    \caption{Adaptive Cache Allocation (ACA)} \label{alg:cache select}
    \textbf{Input:}
    \begin{itemize}
        \item Global class occurrence frequency vector $\bm{\Phi}$.
        \item Timestamp vector $\bm{\tau}^k$ received from client $k$ ($1 \leq k \leq N$).
        \item Hit ratio vector $\mathbf{R}$ and saved inference time vector $\bm{\Upsilon}^k$.
        \item Memory constraint ${\Pi}^k$ for client $k$.
    \end{itemize}
    \textbf{Output:}
    \begin{itemize}
        \item Allocated cache entries indicator matrix $\mathbf{X}^{k}$ of client $k$.
    \end{itemize}
    \begin{algorithmic}[1]
        \For {$i=1$ to $I$ } \label{alg2:1}
            \State Calculate $s^k_i$ by Eq. (\ref{eq:classScore}).
        \EndFor
        \State Sort classes in descending order based on their scores ${s}^k$ to obtain the class order list, $\bm{D}^k$. \label{alg2:3}
        \State $S^k = 0$ \qquad 
        {\color[RGB]{148,0,211}\footnotesize{/* Initialize sum of scores. */}}
        \State $M^k = 0$ \qquad  {\color[RGB]{148,0,211}\footnotesize{/* Initialize sum of size of entries. */}}
        \For {$i$ in $\bm{D}^k$} \label{alg2:6}
                \State Add class $i$ to set $\bm{A}^k$ \ \   {\color[RGB]{148,0,211}\footnotesize{/* Hot-spot classes set $\bm{A}^k$. */}}
            \State $S^k = S^k + s^k_i$
            \If {$S^k \geq \sum_{i=1}^I s^k_i \cdot 0.95$}
                \State Break. \label{alg2:10}
            \EndIf         
        \EndFor
        \While {$M^k \leq {\Pi}^k$ }
            \State $\bm{\zeta}^k = \bm{\Upsilon}^k \cdot \mathbf{R}^k$ {\color[RGB]{148,0,211}\footnotesize{/* Get current expected latency reductions.*/}} \label{alg2:12} 
            \State Find $b$ that maximizes ${\zeta}_{b}^k$
            \State $M^k = M^k + m_j \cdot len(\bm{A}^k)$ {\color[RGB]{148,0,211}\footnotesize{/* $len(\bm{A}^k)$:Length of $\bm{A}^k$.*/}}
            \If {$M^k \ge {\Pi}^k$}
                \State Break.
            \EndIf
            \For {$i$ in $\bm{A}^k$}
                \State $x_{i,b}^k = 1$ \label{alg2:18}
            \EndFor
            \State $p = {R}_b^k$ \label{alg2:19}
            \For {$j=b$ to $L$}
                \State ${R}_j^k = {R}_j^k - p$  \label{alg2:21}
            \EndFor
        \EndWhile
        \State return $\mathbf{X}^k$.   \label{alg2:22}
    \end{algorithmic}
\end{algorithm}

Considering that the class frequency vector and class timestamp vector can represent the frequency and recency of each class, we use these two vectors to estimate the probability of each class reoccurring in the future. Specifically, we calculate a score $s^k_i$ of class $i$ for client $k$ using the following formula:
\begin{equation}
    s^k_i = {\Phi}_i \cdot (0.20)^{ \lfloor \frac{\tau_i^k}{F} \rfloor }
    \label{eq:classScore}
\end{equation}
where ${\Phi}_i$ represents the global frequency of class $i$, while $\tau_i^k$ represents the timestamp for class $i$ on client $k$. We use $F$ to denote the number of inferences in a local inference round. Classes with higher scores have a higher probability of reoccurrence. We then rank the classes based on these scores and select a subset of the top $K$ classes. Following the principles of SMTM, we choose the subset with scores summing up to 95\% of the total score as the hot-spot classes for cache. This ensures a high probability of the class of future inference samples belonging to the selected classes appearing in the cache.

In addition to pursuing a high cache hit ratio and accuracy, we also aim to improve cache efficiency, avoiding excessive lookup time.
Therefore, it is essential to assess the importance of different cache layers and select the most important and effective ones. Vectors $\mathbf{R}$ and $\bm{\Upsilon}$ reflect the expected hit ratio at each cache layer and the saved inference time resulting from cache hits (considering model computation time only). 
We use the dot product $\bm{\zeta}$ of these two vectors to estimate the current expected latency reduction benefit for each cache layer.
Specifically, we iteratively select the cache layer with the highest expected benefit and then recalculate the expected hit ratio and hit benefits for the remaining cache layers until the allocated cache size approaches the constraint threshold. This method ensures that the overhead caused by cache lookup remains within a reasonable range.

Based on the aforementioned design, we propose the two-stage cache entry allocation algorithm ACA (Algorithm \ref{alg:cache select}). The input of ACA include the global class frequency vector $\bm{\Phi}$, the local class timestamp vector $\bm{\tau}^k$, the expected cache hit ratio vector $\mathbf{R}^k$, the saved inference time vector $\bm{\Upsilon}^k$, and the cache size constraint thresholds ${\Pi}^k$. The output is a cache entry allocation indicator matrix $\mathbf{X}^k$.
ACA is divided into two stages:
\begin{enumerate}
    \item In the first stage, ACA calculates importance scores for each class in the cache based on frequency and recency, then sorts them in descending order (Lines \ref{alg2:1}-\ref{alg2:3}). Then, it selects classes accounting for 95\% of the total score as hot-spot classes to form set $\bm{A}^k$ (Lines \ref{alg2:6}-\ref{alg2:10}).
    
    \item In the second stage, ACA evaluates the importance of cache layers and iteratively selects the layer with the highest expected benefit, ensuring the cache size does not exceed the threshold. ACA measures the importance of each cache layer by the product of its expected hit ratio and saved inference time, then selects the layer with the highest benefit. 
    ACA updates the cache size incrementally, stopping just before it exceeds the threshold, and allocates entries in this selected layer according to set $\bm{A}^k$ (Lines \ref{alg2:12}-\ref{alg2:18}). Based on the hypothesis that samples hitting in cache layer $b$ will also hit in cache layer $b+1$, we adjust the expected hit ratio for subsequent layers by subtracting the hit ratio of the current layer (Lines \ref{alg2:19}-\ref{alg2:21}).
\end{enumerate}
Finally, ACA returns the cache entry allocation indicator matrix as output (Line \ref{alg2:22}).

\section{Performance Evaluation}\label{sec:evaluation}
\subsection{Datasets and Models}
\label{subsec:Datasets and Models}

\subsubsection{\textbf{Datasets}}
To validate the effectiveness of our proposed framework CoCa, we employ three real-world datasets for classification and recognition tasks.

\begin{itemize}
    \item \textbf{ImageNet-100 \cite{deng2009imagenet}} is a subset of the ImageNet dataset that contains 100 classes, including animals, plants, common objects, and vehicles. Each class has approximately 1,000 to 1,300 images, totaling around 100,000 images. 
    Our test dataset is organized into batches, with all samples in a batch sharing the same class label to simulate temporal locality. 
    
    \item \textbf{UCF101 \cite{soomro2012ucf101}} is an action recognition dataset with 13,320 videos across 101 action classes, collected from YouTube. It covers five major action classes: human-object interactions, body movements, human-human interactions, playing musical instruments, and sports activities. 
    \item \textbf{Environmental Sound Classification-50 (ESC-50) \cite{piczak2015esc}} is a widely recognized benchmark for environmental sound classification tasks. This curated collection comprises 2,000 labeled environmental audio recordings, each 5 seconds in duration, equally distributed across 50 semantic classes. These classes encompass a diverse range of environmental sounds, including natural soundscapes, human non-speech sounds, and domestic noises.
\end{itemize}

To simulate different data distributions, several methods are employed to construct appropriate datasets for the clients.
Firstly, we construct non-IID data using Dirichlet distribution \cite{wang2023fast}, where $Dir(\epsilon)$ is widely used as the prior distribution in Bayesian statistics, and the parameter $\epsilon$ is adopted to control the non-IID levels. 
A smaller $\epsilon$ means a higher data heterogeneity.
According to \cite{liao2024mergesfl}, we introduce a value $p = \frac{1}{\epsilon}$ to simulate varying non-IID levels.
In our experiments, we choose a set of values $[0,1,2,10]$ of parameter $p$ to represent a range from IID to high non-IID data distributions, with $p=0$ specifically indicating the IID case.
%
Secondly, for the long-tail distribution, we follow the method used in \cite{cao2019learning}. 
We denote the sample number of class $i$ as ${d}_i$. The imbalance ratio $\rho$ is defined as the ratio between the sample sizes of the most frequent and least frequent classes. Specifically, $\rho = {\max_{1 \leq i \leq I} {d}_i }/{ \min_{1 \leq j \leq I} {d}_j }$, where $I$ is the total number of classes and $\rho \geq 1$.
The constructed long-tail distribution follows an exponential decay in sample sizes across different classes. 
To create a dataset with 100 classes from ImageNet-100 where the top 20\% most frequent classes possess approximately 60\% of the total samples, we set ${\rho} = 90$.

\subsubsection{\textbf{Models}}
To comprehensively assess the performance of CoCa across various types and depths of neural networks, we employ five distinct models on the aforementioned datasets. 

\begin{itemize}
    \item \textbf{VGG16\_BN \cite{sengupta2019going}} consists of 13 convolutional layers using 3$\times$3 convolutional kernels, 2 dense layers, and a softmax output layer. Additionally, the model incorporates batch normalization layers to optimize the training process.
    
    \item \textbf{ResNet50 \cite{he2016deep}} is composed of a 7$\times$7 convolutional layer, 16 residual blocks, and a fully connected classify layer. Each residual block consists of 3 different convolutional layers, featuring residual connections that skip several layers to alleviate the vanishing gradient problem. The 16 residual blocks are distributed across four sections in the configuration of 3, 4, 6, and 3 blocks.
    
    \item \textbf{ResNet101 \cite{lin2021application}} shares a similar backbone structure as ResNet50, but with an increased depth of 101 layers. 
    The 33 residual blocks are distributed across four sections in the configuration of 3, 4, 23, and 3 blocks.
    
    \item \textbf{ResNet152 \cite{prabhakaran2021thermal}} boasts an astonishing depth of 152 layers. The 50 residual blocks are distributed across four sections in the configuration of 3, 8, 36, and 3 blocks.

    \item \textbf{Audio Spectrogram Transformer (AST) \cite{gong2021ast}} is a deep learning model designed for audio classification tasks. AST adapts the Vision Transformer (ViT) architecture to the audio domain, processing spectrograms directly without manual feature extraction. We choose the AST-Base with 12 transformer block.
\end{itemize}

\subsection{Baselines and Metrics}

\subsubsection{\textbf{Baselines}}
To comprehensively evaluate the performance of CoCa, we choose the following four baselines:

\begin{itemize}
    \item \textbf{Edge-Only} represents the conventional edge inference approach where clients perform inference solely on their devices. This baseline helps establish the standard performance metrics without additional acceleration techniques, providing a reference for evaluating the effectiveness of advanced methods.
    
    \item \textbf{LearnedCache} \cite{balasubramanian2021accelerating} is designed to ensure stable memory usage and consistent time consumption by setting multiple intermediate exits in the model. It allows samples to exit early to save inference time. 
    The decision to exit early or not is determined by a separate small model at each exit.
    We introduce this baseline to compare how model-based multi-exit strategies perform against caching-enhanced methods.
    
    \item \textbf{FoggyCache} \cite{guo2018foggycache} demonstrates the potential of client-server frameworks in leveraging caching for inference acceleration. Different from the semantic caching mechanism, FoggyCache uses the modified A-LSH algorithm for organizing cache objects and uses H-kNN for cache lookup before inference. 
    It highlights the effectiveness of server-assisted caching in reducing inference latency. 
    This baseline serves as a comparison object for a different single-exit caching framework in multi-client scenarios.
    
    \item \textbf{SMTM} \cite{li2021boosting} implements multi-exit inference acceleration using a semantic caching mechanism. It reduces cache space occupancy through compressed cache entry representations. We extend its application to multiple clients as a baseline for multi-exit caching in multi-client scenarios.
\end{itemize}

\subsubsection{\textbf{Metrics}}
We adopt the following two metrics to evaluate the performance of CoCa and the baselines.

\begin{itemize}
    \item \textbf{Average Latency (ms).} We measure the inference latency for each sample on each client's dataset. The average inference latency is calculated as the total inference time divided by the total number of samples across all clients. 

    \item \textbf{Overall Accuracy (\%).} We define the overall accuracy, which calculates the proportion of correctly inference samples on all clients to the total sample, to evaluate the inference performance of the various acceleration methods. 
\end{itemize}


\subsection{System Implementation}
We utilize a simulated edge computing environment based on the NVIDIA Jetson TX2 \cite{suzen2020benchmark} as our experimental platform.
The Jetson devices are configured as clients and servers responsible for loading inference models and datasets. The experimental network is established using a router, enabling wireless connections between clients and the server. Our software implementation employs Docker Swarm \cite{singh2023load} for distributed software development and the PyTorch deep learning framework. Docker Swarm facilitates the development of distributed solutions and provides monitoring capabilities for each client's status. The PyTorch deep learning framework enables the efficient execution of complex inference operations on clients. Moreover, the communication system of the experimental platform is built using Message Passing Interface (MPI) \cite{ragunthar2021strong}, optimizing parallel communication functions through send and receive operations. 


\subsection{Threshold Configuration}
\begin{figure}[t]
\subfigure[ VGG16\_BN.]{
\label{theta_effect.a}
\centering
\includegraphics[width=1.6in]{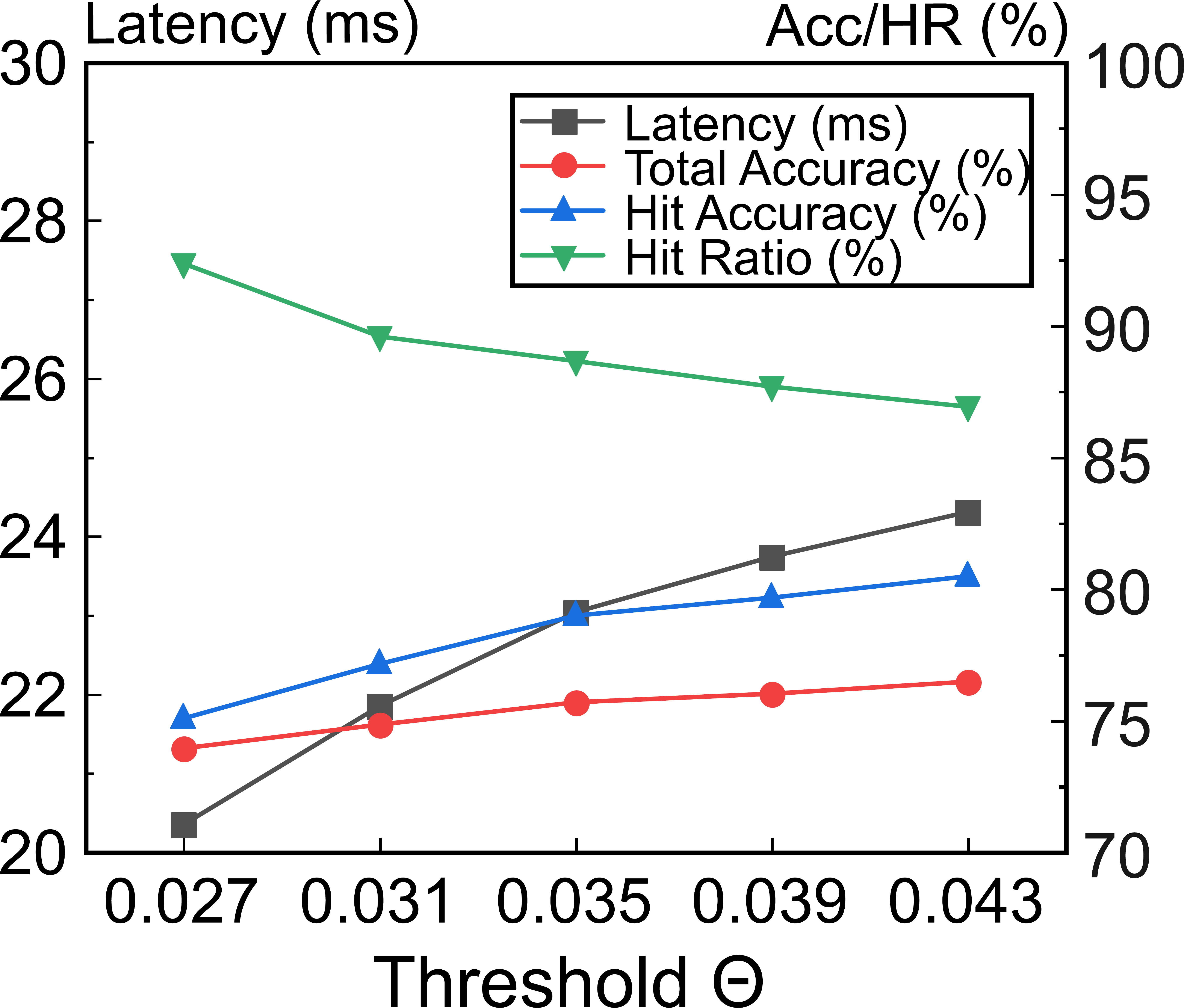}
}
\subfigure[ResNet101.]{
\label{theta_effect.b}
\centering
\includegraphics[width=1.6in]{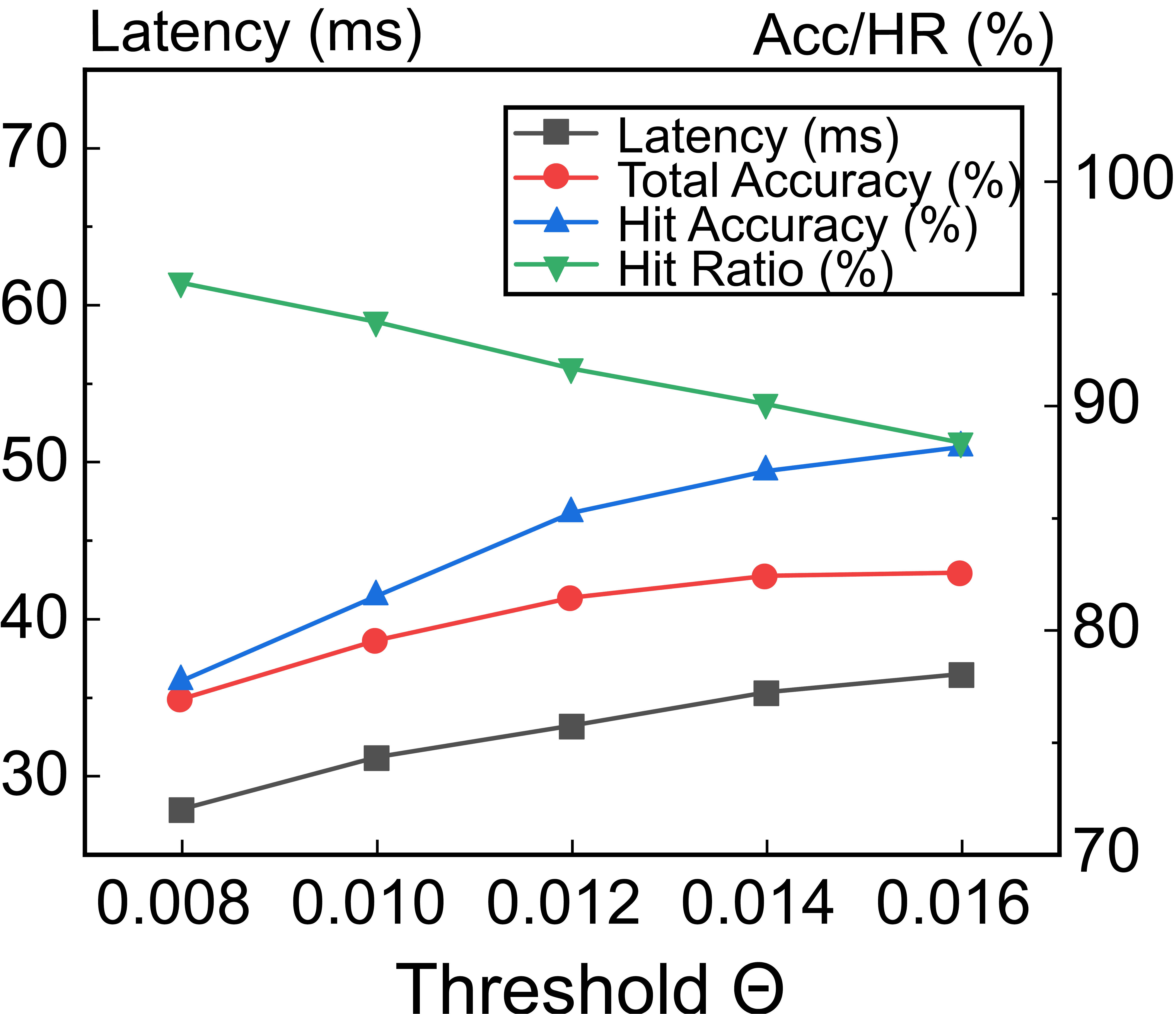}
}
\caption{The impact of different threshold values $\Theta$. }\label{theta_effect}\vspace{-5mm}
\end{figure}

In the CoCa system, we determine the threshold $\Theta$ based on empirical information obtained from comprehensive testing and the desired service-level objectives (such as accuracy loss constraints). 
We evaluate the impact of different threshold values $\Theta$ on inference performance, such as cache hit ratio, cache hit accuracy, average inference latency, and overall inference accuracy, across different models and datasets.
The test results are illustrated in Fig. \ref{theta_effect}.
As the threshold $\Theta$ increases from 0.008 to 0.016, the cache hit ratio of ResNet101 declines from 95.5\% to 88.3\% due to the stricter cache hit criteria, which reduces the probability of a hit. 
However, as $\Theta$ increases, cache hit accuracy, overall inference accuracy, and average inference latency all improve. For example, the cache hit accuracy of VGG16\_BN continues to improve when $\Theta$ rises. After $\Theta$ reaches 0.035, the rate of improvement begins to decelerate.
Selecting the optimal $\Theta$ requires balancing SLOs with performance gains.
%
We observe that the optimal threshold $\Theta$ remains similar across different datasets (\eg, ImageNet100 and UCF101) for the same model. 
Based on our experimental results, we select the appropriate threshold $\Theta$ for the ResNet models under different accuracy loss constraints. Specifically, we set $\Theta = 0.012$ to meet the 3\% accuracy loss requirement, and $\Theta = 0.008$ to satisfy the 5\% accuracy loss constraint.
For VGG16\_BN, the corresponding recommended thresholds are 0.035 and 0.027, respectively.

\begin{figure}[t]
\subfigure[Absorption ratio/accuracy vs. $\Gamma$.]{
\label{Gamma_Effect}
\centering
\includegraphics[width=1.6in]{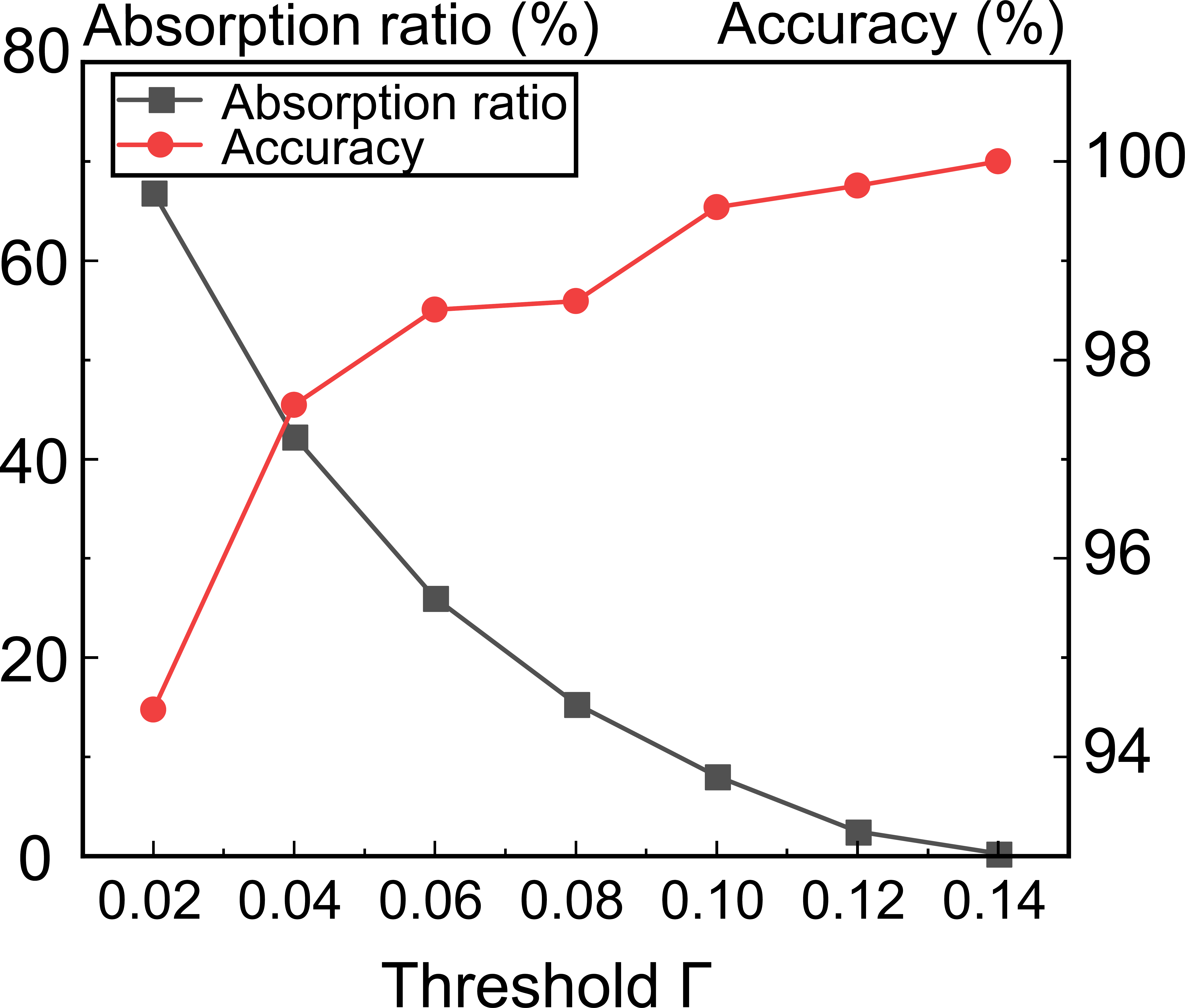}
}
\subfigure[Absorption ratio/accuracy vs. $\Delta$.]{
\label{Delta_Effect}
\centering
\includegraphics[width=1.6in]{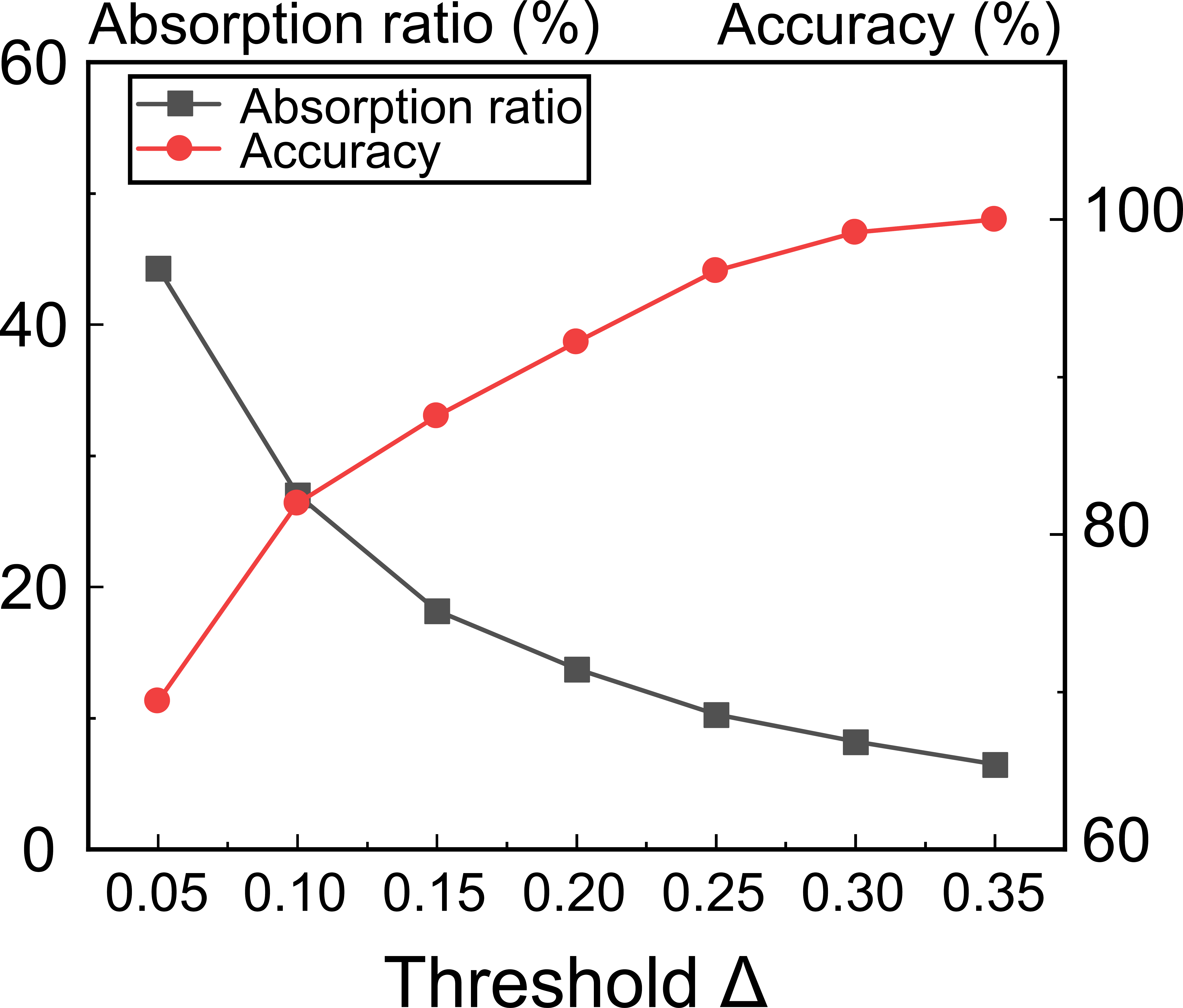}
}
\caption{The impact of different threshold values $\Gamma$ and $\Delta$. }
\label{gamma_delta}
\end{figure}

The thresholds, $\Gamma$ and $\Delta$, used for sample selection, play a crucial role in determining the collection of two types of inference samples (\ie, cache hits and cache misses) for global cache updates. 
We test different threshold configurations and measure the absorption ratio (\ie, the probability that a sample is collected to update the global cache after satisfying the preconditions) and accuracy for both types of samples on UCF101 with ResNet101, as shown in Fig. \ref{gamma_delta}. 
As the thresholds increase, the absorption ratio for both types of samples gradually decreases, while the accuracy steadily improves. 
When $\Gamma = 0.14$ and $\Delta = 0.35$, the accuracy of both sample types can even reach 100\%, but the absorption ratios drop to 0.21\% and 6.47\%, respectively. 
This is because smaller thresholds lead to more precise sample selection. 
The selection of thresholds must balance the reinforcement of the current vector center and the exploration of new semantic vectors. 
Based on our experiments, we recommend thresholds that achieve over 97\% accuracy with an absorption ratio around 10\%, striking a balance between global cache update effectiveness and the cost of sample collection. For example, we suggest $\Gamma = 0.1$ and $\Delta = 0.25$ for the ResNet models.

\subsection{Latency under SLOs}

\begin{table}[t]
\centering
\setlength{\abovecaptionskip}{10pt}%
\setlength{\belowcaptionskip}{0pt}%
\footnotesize
\caption{Inference latency (Lat./ms) and accuracy (Acc./\%) on a 100-class subset of the UCF101 dataset within given accuracy loss constraint. Edge-Only provides the standard latency and accuracy performance used as constraint benchmarks.}
\label{table:Acc cons}
\begin{tabular}{ c c c c c c }
\hline
\rule{0pt}{8pt}
\multirow{2}{*}{ Models } & \multirow{2}{*}{ Methods } & \multicolumn{2}{c}{<3\%} & \multicolumn{2}{c}{<5\%} \\
\cline{3-6}
\rule{0pt}{8pt}
 &  & Lat. & Acc. & Lat. & Acc. \\
\hline
\rule{0pt}{8pt}
\multirow{5}{*}{VGG16\_BN} & Edge-Only           &  29.94 & 78.12 & 29.94 & 78.12   \\
\cline{2-6}
\rule{0pt}{8pt}
& LearnedCache & 26.82 & 75.28 & 23.98 & 73.86 \\
\cline{2-6}
\rule{0pt}{8pt}
& FoggyCache & 27.12 & 75.37 & 23.42 & 73.89 \\
\cline{2-6}
\rule{0pt}{8pt}
& SMTM & 25.51  & 75.41 & 22.15  & 73.97   \\
\cline{2-6}
\rule{0pt}{8pt}
& \textbf{CoCa} & \textbf{23.05} & {75.73} & \textbf{20.36} & {74.23}  \\
\hline
\rule{0pt}{8pt}
\multirow{5}{*}{ResNet152} & Edge-Only              & 62.85 & 83.98 & 62.85 & 83.98 \\
\cline{2-6}
\rule{0pt}{8pt}
& LearnedCache & 48.39 & 81.30 & 43.74 & 79.31 \\
\cline{2-6}
\rule{0pt}{8pt}
& FoggyCache & 45.22 & 81.63 & 39.69 & 79.51 \\
\cline{2-6}
\rule{0pt}{8pt}
& SMTM & 44.83  & 81.72 & 35.54  & 79.46   \\
\cline{2-6}
\rule{0pt}{8pt}
& \textbf{CoCa} & \textbf{34.45} & {82.16} & \textbf{29.53} & {79.82}  \\
\hline
\end{tabular}
\end{table}


%
We conduct experiments on the UCF101 datasets to evaluate the average inference latency of CoCa and various baselines with specific accuracy loss constraint SLOs (\eg, 3\% and 5\% \cite{xu2017accelerating}). 
Table \ref{table:Acc cons} presents the test results on a subset of 100 classes from UCF101.
CoCa consistently achieves greater latency reductions within the given accuracy loss constraints compared to the baselines. 
For example, under an accuracy loss constraint of 3\%, CoCa reduces the average inference latency by 23.01\% and 45.19\% on VGG16\_BN and ResNet152 respectively, compared to Edge-Only. 
Additionally,  CoCa achieves a latency reduction of 23.2\% to 28.8\% on ResNet125 compared to other acceleration baselines.
Unlike traditional multi-exit methods, CoCa employs semantic caching to reduce computational overhead, resulting in improved latency reductions. Compared to SMTM, CoCa utilizes a more granular dynamic local cache allocation, which better balances accuracy and latency performance, and achieves lower average inference latency.
In summary, CoCa demonstrates significant improvements in inference latency across different models and datasets, showing its potential as an effective acceleration framework.

\subsection{The Effect of Data Distribution}

\subsubsection{\textbf{Non-IID distribution}}

\begin{figure}[t]\centering
    \includegraphics[width=0.48\textwidth]{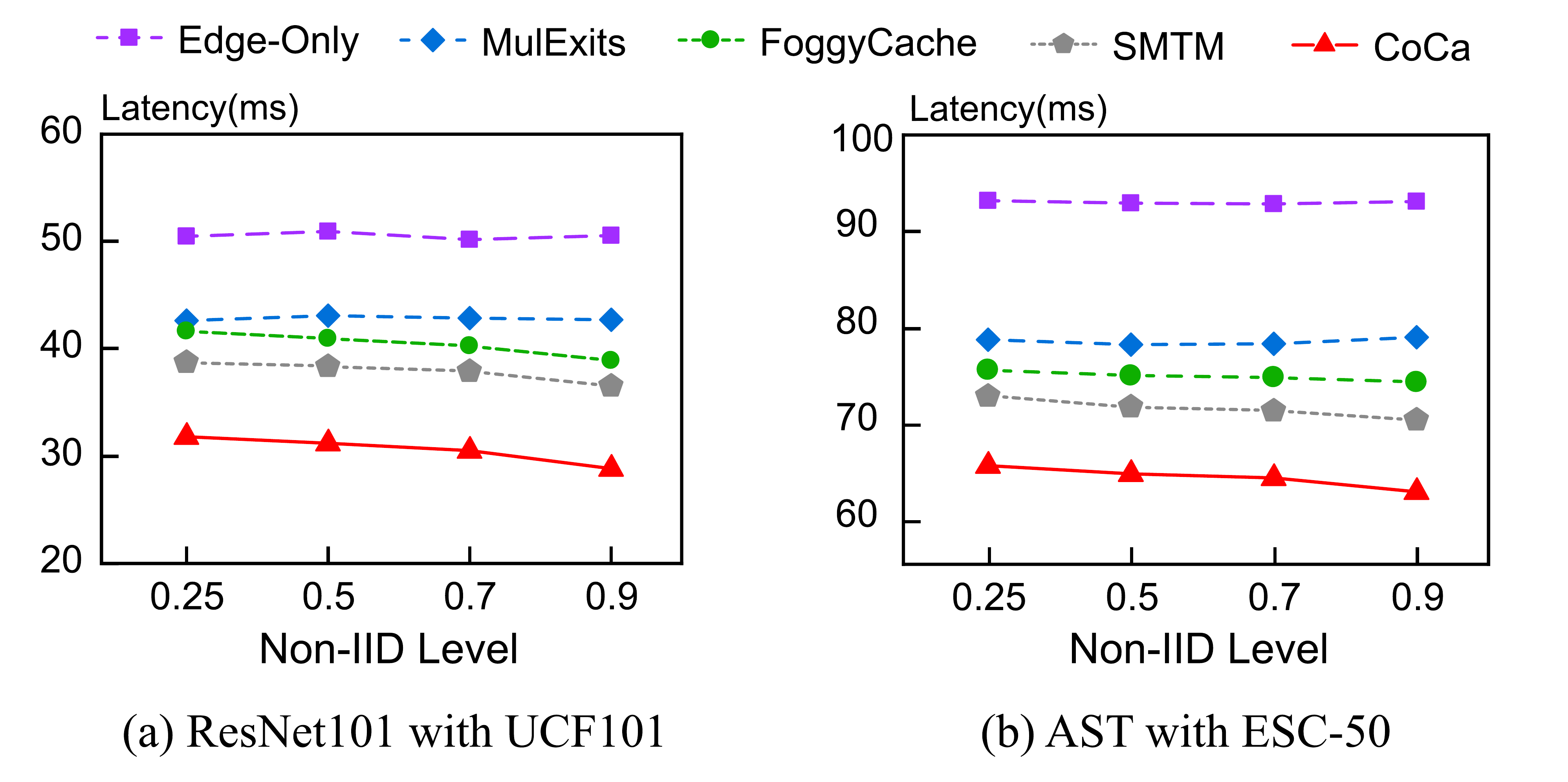}
    \caption{The latency (ms) under different non-IID levels. }\label{Non-IID_Latency}\vspace{-3mm}
\end{figure}

To assess the performance of CoCa on non-IID data, we conduct experiments on ESC-50 and a subset of 100 classes from UCF101 with varying non-IID levels ($p=0, 1, 2, 10$), and the test results are shown in Fig. \ref{Non-IID_Latency}.
Methods without caching show stable latency across different non-IID levels. In contrast, caching-based methods exhibit decreasing latency as non-IID levels increase. 
The average inference latency on two models of CoCa consistently decreases with increasing non-IID level. Under the highest non-IID level ($p=10$), ResNet101's average latency decreases by 9.08\% compared to the IID case.
CoCa consistently outperforms other baselines in latency decrease across different models and non-IID levels. For AST, CoCa achieves a 29.02\% to 33.17\% decrease in latency compared to Edge-Only and a 12.19\% to 15.43\% decrease compared to SMTM.
As non-IID level increases, temporal locality becomes more pronounced, enhancing the effectiveness of cache-based methods. CoCa's superior performance is primarily due to global updates, which improve both the cache hit ratio and inference accuracy.
In summary, CoCa significantly reduces average inference latency while maintaining higher inference accuracy compared to other acceleration baselines, demonstrating its robustness and effectiveness, especially with highly non-IID data distributions.

\subsubsection{\textbf{Long-tail distribution}}
For the long-tail distribution, we establish two groups: one with a uniform distribution across classes (uniform group) and the other following a long-tail distribution (long-tail group), based on a subset of 100 classes from ImageNet-100.
As described in Section \ref{subsec:Datasets and Models}, the dataset of the long-tail group exhibits a long-tail distribution where approximately 60\% of the data falls within the top 20\% most frequent classes.
The inference latency and accuracy in these two groups are shown in Table \ref{table:longtail}. 
LearnedCache and FoggyCache perform similarly across uniform and long-tail groups. CoCa and SMTM exhibit significantly lower inference latency in the long-tail group compared to the uniform group. 
CoCa outperforms all baselines in both groups, achieving the lowest latency while maintaining competitive accuracy. Specially, CoCa reduces the inference latency by 4.01\% in the long-tail group compared to the uniform group, while maintaining an accuracy of 81.35\%.
This can be attributed to two factors. 
Firstly, the higher proportion of frequent classes in the long-tail group indirectly enhances temporal locality within the dataset, thereby resulting in an increased cache hit ratio. 
Secondly, the cache allocation strategy of CoCa handles long-tail distributions more effectively compared to the simple LRU strategy.
In summary, by adapting to the data distribution and optimizing cache allocation, CoCa significantly reduces latency while maintaining accuracy, which suggests the potential benefits of CoCa in real-world scenarios with long-tail data distributions.



\begin{table}[t]
\centering
\setlength{\abovecaptionskip}{10pt}%
\setlength{\belowcaptionskip}{0pt}%
\footnotesize
\caption{The average inference latency (Lat./ms) and accuracy (Acc./\%) of ResNet101 on constructed uniformed and long-tail groups dataset of ImageNet-100.}
\label{table:longtail}
\begin{tabular}{ c c c c c}
\hline
\rule{0pt}{8pt}
            & \multicolumn{2}{c}{Uniformed} &  \multicolumn{2}{c}{Long-tail} \\
\cline{2-5}
\rule{0pt}{8pt}
            & Lat. & Acc. & Lat. & Acc. \\
\hline
\rule{0pt}{8pt}
Edge-Only      & 44.87 & 82.64  & 44.91  &  82.67   \\
\hline
\rule{0pt}{8pt}
LearnedCache    & 35.81 & 80.56  & 36.02  &  80.61   \\
\hline
\rule{0pt}{8pt}
FoggyCache  & 34.64 & 80.83  & 34.83  & 80.88    \\
\hline
\rule{0pt}{8pt}
SMTM        & 34.02 & 80.68  & 33.29  & 80.92    \\
\hline
\rule{0pt}{8pt}
\textbf{CoCa}        & \textbf{28.17} & {80.97}  & \textbf{27.04}  & {81.35}    \\
\hline
\end{tabular}
\end{table}

\subsection{ACA Performance}\label{ablation}
\begin{figure}[t]\centering
    \includegraphics[width=0.3\textwidth]{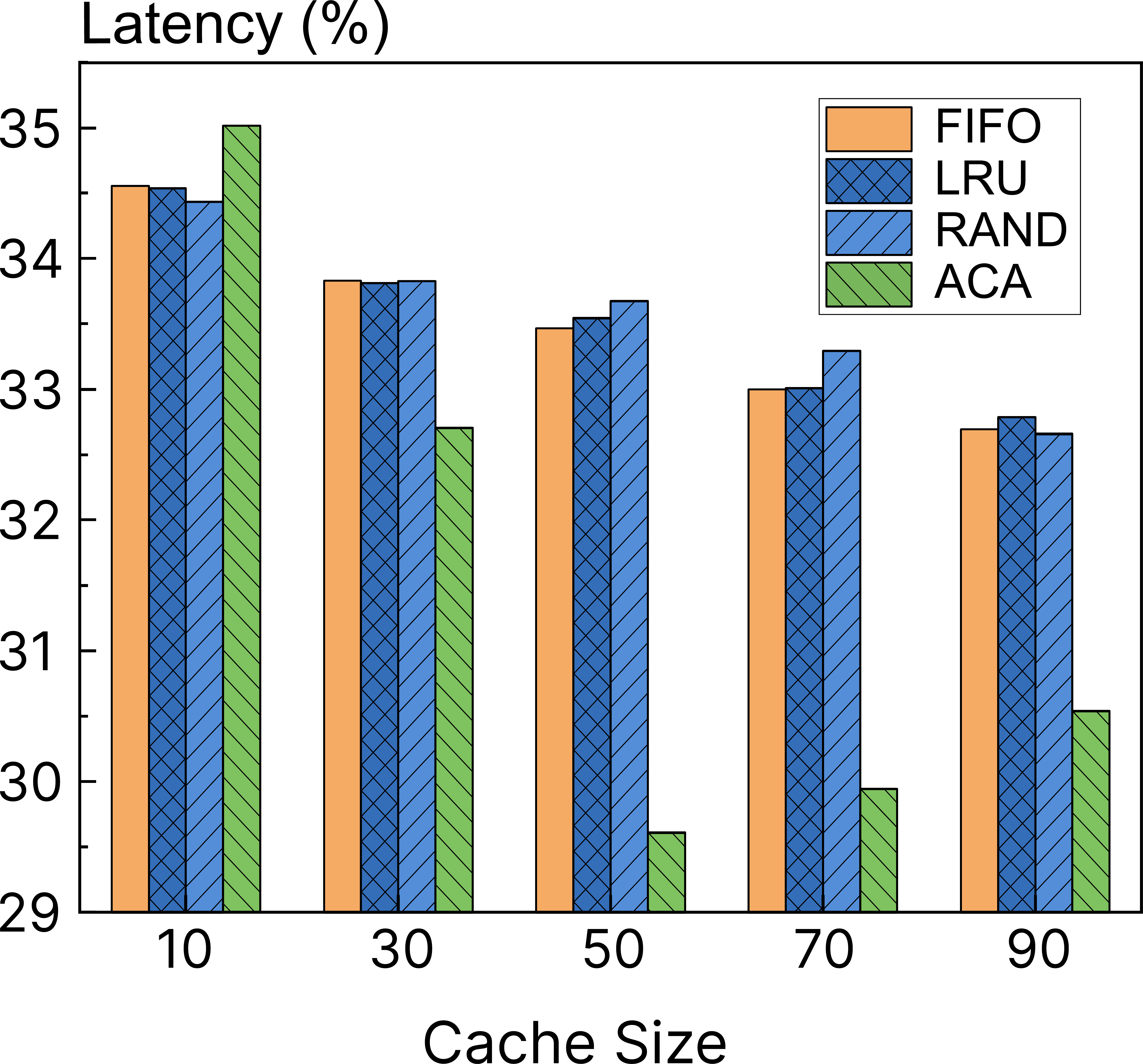}
    \caption{Performance comparison of different cache replacement algorithms.}\label{fig_diff_cr_strategy}\vspace{-0.5cm}
\end{figure}

To demonstrate the effectiveness of the ACA algorithm, we compare it with several classical cache replacement algorithms, including LRU, FIFO, and RAND.  
Specifically, LRU assigns a counter to each cache entry to track how long it has not been accessed and replaces the entry with the largest counter value during replacement.
FIFO replaces the earliest loaded cache entry when the cache is full, while RAND randomly selects an entry for replacement when the cache reaches its capacity.
For the baseline algorithms, we select a fixed set of cache layers with high expected benefit (define in \ref{sec:algorithm}) to serve as local cache. We define cache size as the maximum number of entries each cache layer can accommodate. 
To ensure a fair comparison, ACA is constrained to use the same total memory size as the baseline algorithms. 
We conduct experiments to evaluate the latency performance of ACA and baseline algorithms using a long-tail distributed UCF101 dataset containing 100 classes. These experiments are performed with varying cache sizes.
Specifically, we use the average latency metric under a 3\% accuracy loss constraint as an example and the results of our comparative analysis are presented in Fig. \ref{fig_diff_cr_strategy}.

It can be observed that the inference performance (\ie, latency) of all methods decreases as the cache size increases. 
When the cache size exceeds 30, our proposed algorithm significantly outperforms the various baseline strategies.
This is because, with a small number of cache entries, ACA selects fewer cache layers to ensure coverage, resulting in limited improvements. 
As the cache size increases, the number of selected cache layers also increases, leading to a more significant reduction in inference latency. 
However, as the cache size continues to grow, the gains from cache hits become smaller compared to the additional lookup overhead, causing the average inference latency to rise again. 
In conclusion, the experimental results can significantly demonstrate the correctness and excellent performance of the ACA algorithm.

\subsection{Ablation Study}\label{ablation}

\begin{figure}[t]
  \vspace{1.9mm}\centering
  \subfigure[Latency]{\includegraphics[width=1.7in]{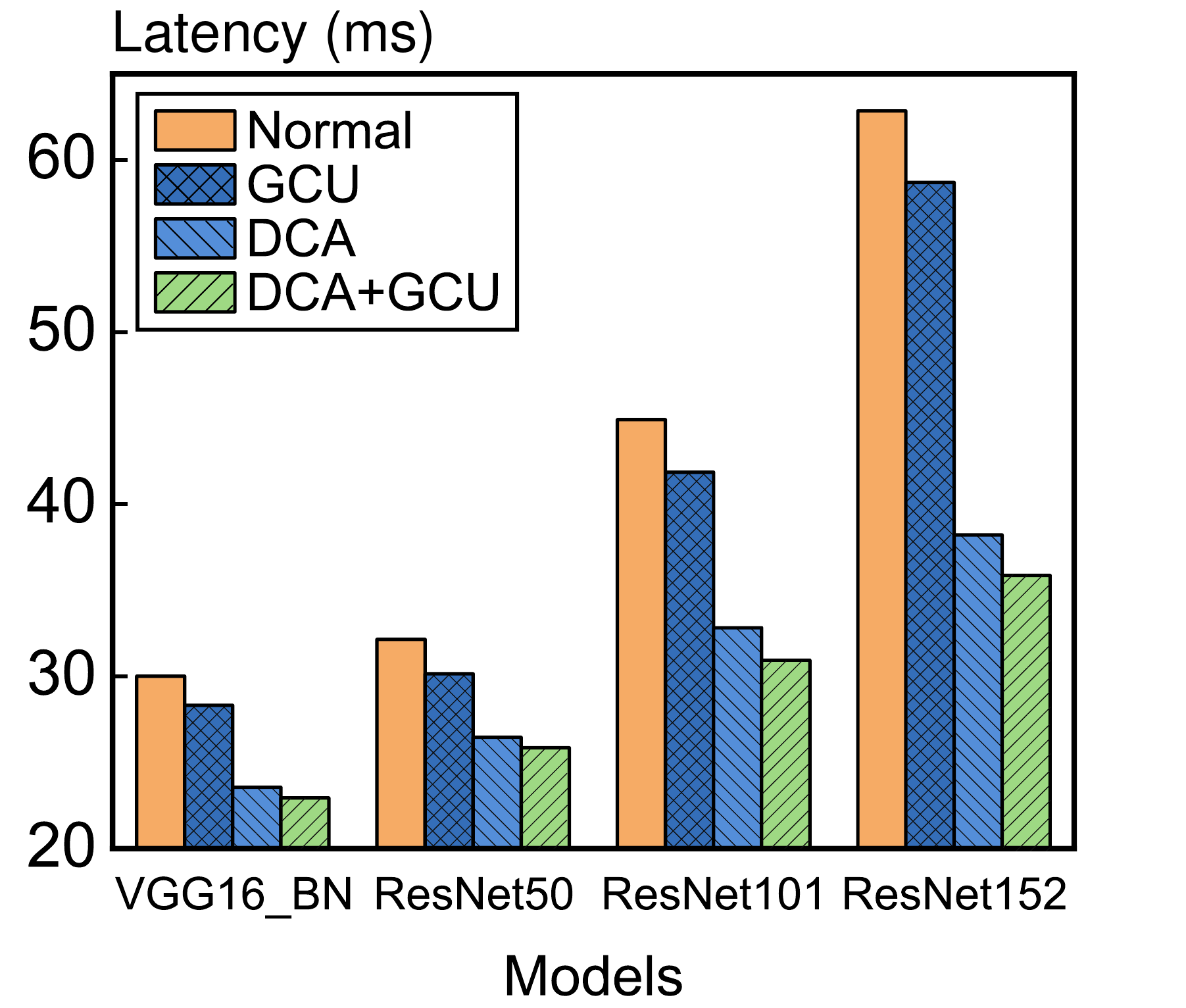}}
  \subfigure[Accuracy]{\includegraphics[width=1.7in]{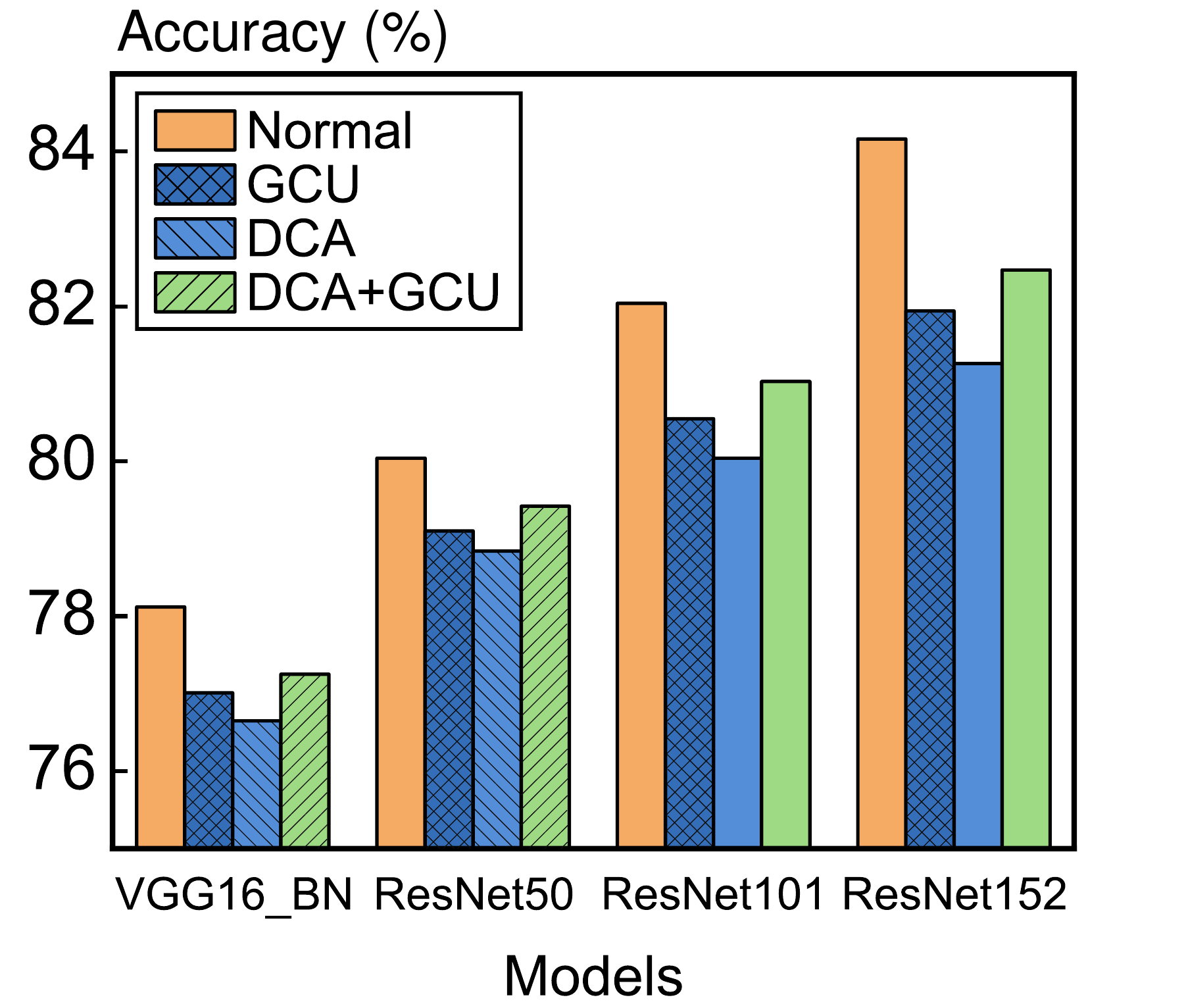}}
  \caption{Latency and accuracy performance on 50-class subset of the UCF101 dataset across different models.}
  \label{fig:Ablation}
\end{figure}

To investigate the specific effects of two main components of CoCa, \ie, dynamic cache allocation (DCA), and global cache updates (GCU), on average inference latency and accuracy, we conduct the following ablation experiments. DCA dynamically adjusts the number of cache layers and hot-spot classes during inference, in contrast to fixed cache configurations. 
GCU updates the global cache table throughout the inference process, as opposed to using a static global cache.
Four approaches are compared in terms of their performance: Normal (standard), DCA (dynamic cache allocation only), GCU (global cache updates only), and DCA+GCU (utilizing both two components).
The results in Fig. \ref{fig:Ablation} indicate that employing both improvements simultaneously (\ie, DCA+GCU) yields superior performance in terms of average inference latency and accuracy compared to utilizing either improvement alone (DCA or GCU).
The DCA can achieve a more significant latency reduction across all four models compared to the GCU. 
For instance, we compare the performance of different methods on ResNet152, using Normal as the baseline. The results show that GCU achieves a reduction in latency of 6.60\%. In contrast, DCA demonstrates superior performance with a remarkable 39.18\% reduction in latency. 
In other words, DCA outperforms GCU by a significant margin of 32.58\% in latency reduction, suggesting that DCA plays a dominant role in latency reduction, as it directly impacts cache lookup overhead and hit ratio. In contrast, GCU indirectly improves the cache hit ratio by using more representative cache entries, resulting in a relatively smaller impact.
In terms of accuracy performance, GCU exhibits relatively lower accuracy loss across all four models compared to DCA. For ResNet101, the accuracy loss of DCA relative to Normal is 2.01\%, while GCU achieves an accuracy loss of only 1.49\%, a reduction of 0.52\%, or nearly 25\%. 
The combination of both components (\ie, DCA+GCU) results in the best accuracy performance. This is likely due to the combinational effects where GCU enhances the representativeness of cache entries for DCA, while DCA provides better cache management to avoid excessive interference from redundant cache entries.
The above findings demonstrate the complementary nature of the two key components of CoCa and the importance of their joint application to achieve optimal performance in terms of both latency reduction and accuracy maintenance.

\subsection{System Load Analysis}

\begin{figure}[t]
  \centering
  \subfigure[Accuracy/latency vs. $F$.]{  \label{numFrames}\includegraphics[width=1.7in]{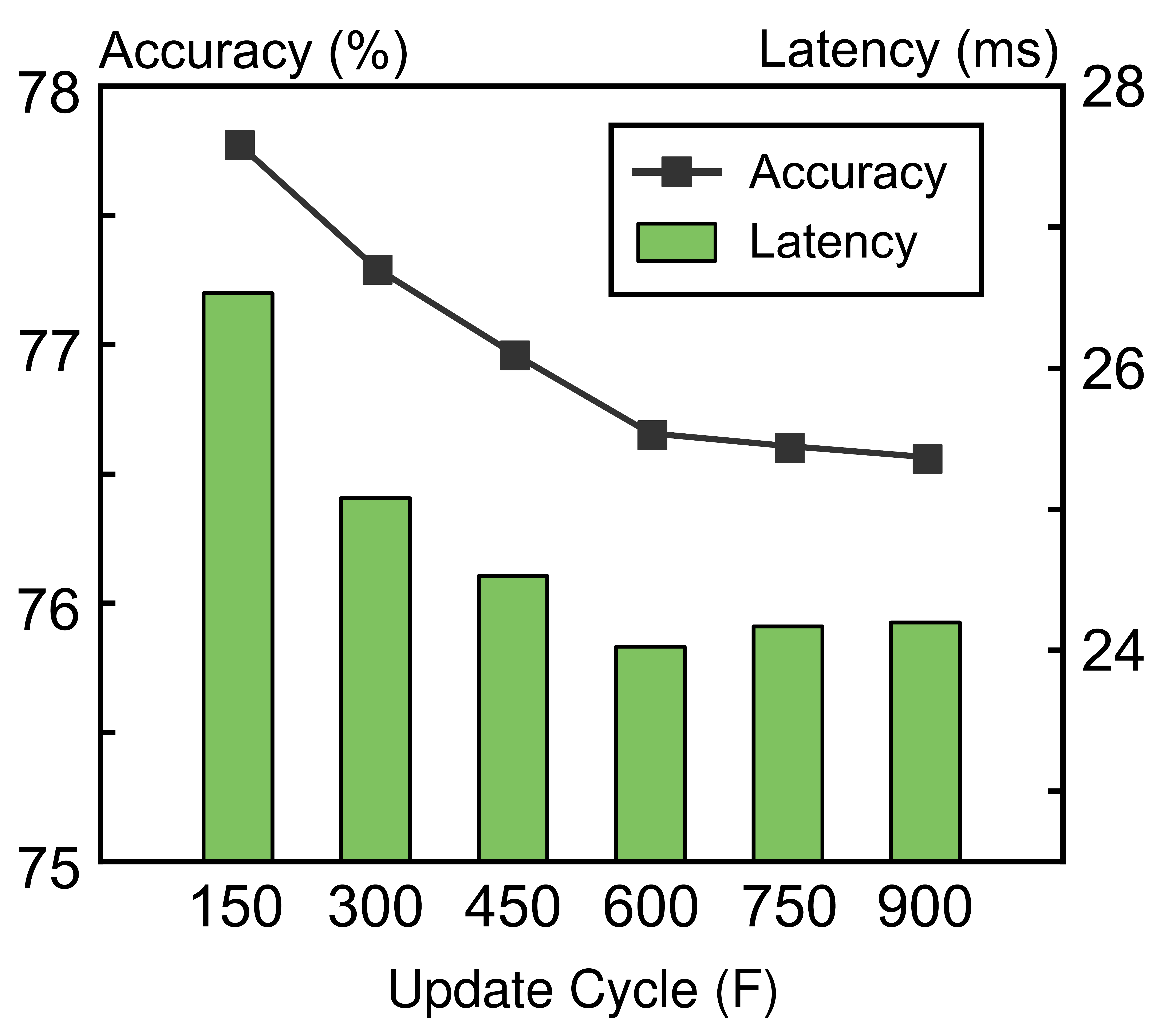}}
  \subfigure[Accuracy vs. number of clients.]{\label{numclients}\includegraphics[width=1.7in]{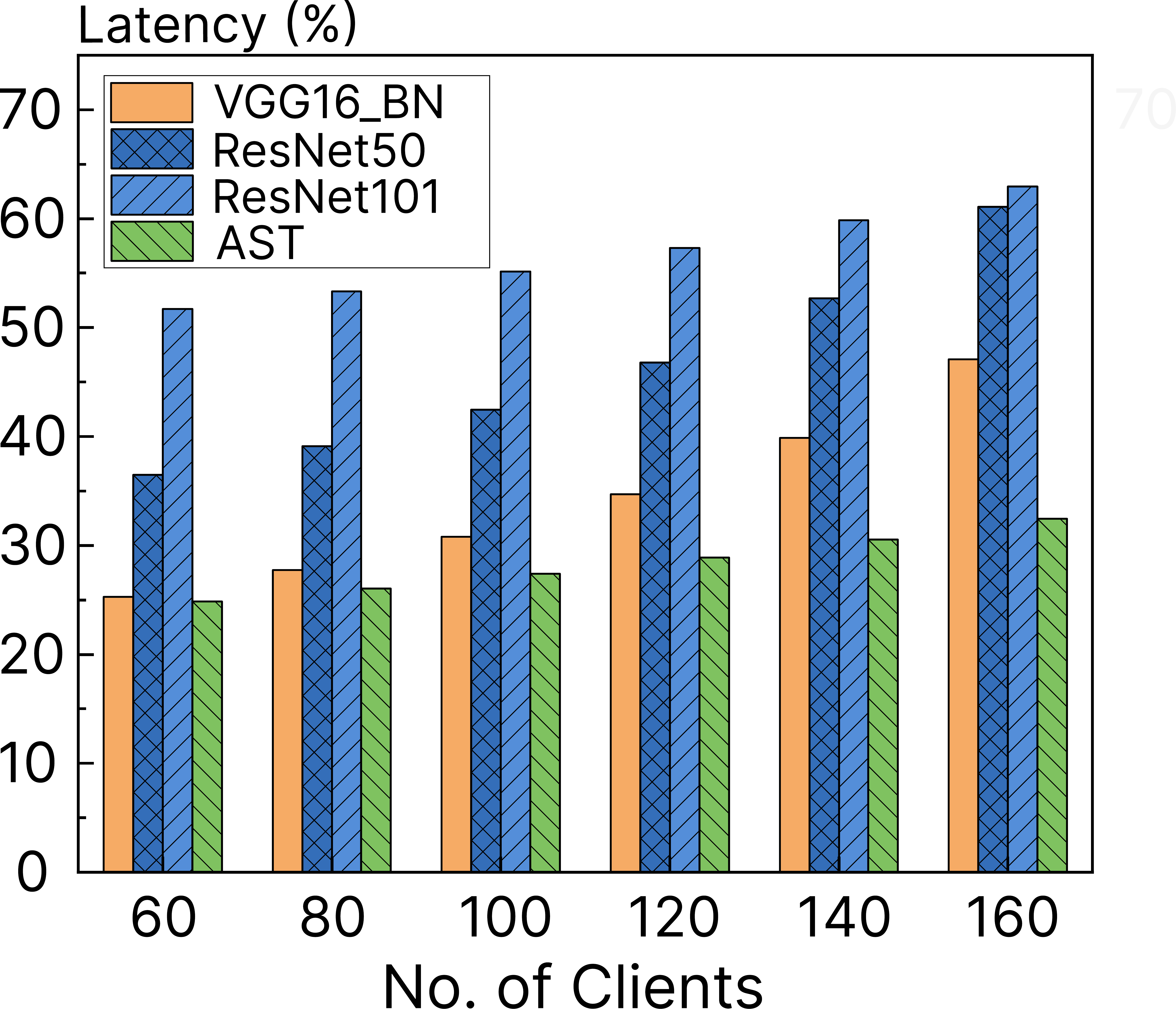}}
  \caption{ The effect of update cycles and the number of clients.}
  \label{fig:Scale}
\end{figure}

%

Since the clients need to frequently request cache allocation from the server in CoCa, we conduct system load testing and analyze the impact of varying cache update cycles and the number of clients on latency and accuracy.
The results are shown in Fig. \ref{fig:Scale}. 
We adopt $F$ to represent the update cycle, \ie, the number of inference frames for a cache update.
Fig. \ref{numFrames} illustrates the average latency and accuracy performance of VGG16\_BN on a long-tailed dataset of 100 classes from UCF101 with different values of $F$.
As the cycle $F$ increases, average inference latency initially decreases significantly before stabilizing when $F \ge 300$. The latency reduces from 26.54ms ($F=150$) to 24.02ms ($F=900$). Inference accuracy exhibits a similar declining trend.
Given lower values of $F$, the likelihood of simultaneous client requests increases, leading to increased waiting times and server-side overhead. 
As $F$ increases, the load of the server decreases, but cache timeliness decreases, potentially affecting accuracy.
Considering the trade-off between accuracy loss and latency reduction, we set $F = 300$.

Fig. \ref{numclients} illustrates that increasing the number of clients leads to a slight increase in the average response latency for cache requests (defined as the time elapsed between issuing a request and receiving the cached data). 
For instance, the average response latency for ResNet101 rises from 56.70ms (60 clients) to 60.93ms (160 clients), representing a modest increase of 7.46\%.
This slight degradation in average response latency performance is attributed to increased competition for global cache access as the number of clients grows, which also increases the overhead on the server.
Nevertheless, in our proposed method, the exchanged cache size between client and server remains typically small ($<$ 1 MB), keeping the server load minimal and supporting strong system scalability even under high request volumes. 
Based on the analysis of the impact of the update cycles and the number of clients on inference performance, our proposed CoCa can achieve good scalability.

\section{Conclusion}\label{sec:conclusion}
In this paper, we propose an efficient inference framework CoCa, which adopts a multi-client collaborative caching mechanism, to accelerate edge inference.
Specifically, CoCa leverages two core mechanisms, \ie, dynamic cache allocation and global cache updates,
to address the challenges of long-tail and non-IID distribution. 
We evaluate the inference performance of CoCa with three classical datasets and five distinct models respectively.
Extensive experiments demonstrate that CoCa reduces inference latency by 23.01\% to 45.19\% with a slight ($<$ 3\%) accuracy loss. 
These experimental results confirm the effectiveness and efficiency of CoCa.


\bibliographystyle{IEEEtran}
\bibliography{content/refs} 


\end{document}